\documentclass[final,3p,times,twocolumn]{elsarticle}

\usepackage{graphicx}

\usepackage{amssymb}

\usepackage{amsmath} 
\usepackage{epstopdf}
\usepackage{booktabs}

\journal{Nuclear Instruments and Methods A}

\begin{document}

\begin{frontmatter}



\title{An apparatus for studying spallation neutrons in the Aberdeen Tunnel laboratory}


\author[nuu]{S. C. Blyth}
\author[cuhk]{Y. L. Chan}
\author[cuhk]{X. C. Chen}
\author[cuhk]{M. C. Chu}
\author[bnl]{R. L. Hahn}
\author[ntu]{T. H. Ho}
\author[ntu]{Y. B. Hsiung}
\author[nctu]{B. Z. Hu}
\author[cuhk]{K. K. Kwan}
\author[cuhk]{M. W. Kwok}
\author[hku]{T. Kwok}\ead{tnkwok@hku.hk}
\author[hku]{Y. P. Lau}
\author[hku]{K. P. Lee}
\author[hku]{J. K. C. Leung}
\author[hku]{K. Y. Leung}
\author[nctu]{G. L. Lin}
\author[cuhk]{Y. C. Lin} 
\author[ucb]{K. B. Luk}
\author[cuhk]{W. H. Luk}
\author[hku]{H. Y. Ngai}
\author[cuhk]{S. Y. Ngan}
\author[hku]{C. S. J. Pun}
\author[cuhk]{K. Shih}
\author[cuhk]{Y. H. Tam}
\author[hku]{R. H. M. Tsang}
\author[nuu]{C. H. Wang}
\author[cuhk]{C. M. Wong}
\author[hku]{H. L. Wong}
\author[hku]{H. H. C. Wong}
\author[cuhk]{K. K. Wong}
\author[bnl]{M. Yeh}

\address[bnl]{Chemistry Department, Brookhaven National Laboratory, Upton, NY 11973, USA}
\address[nuu]{Department of Electro-Optical Engineering, National United University, Miao-Li, Taiwan}
\address[ntu]{Department of Physics, National Taiwan University, Taipei, Taiwan}
\address[cuhk]{Department of Physics, Chinese University of Hong Kong, Hong Kong, China}
\address[hku]{Department of Physics, University of Hong Kong, Hong Kong, China}
\address[ucb]{Department of Physics, University of California at Berkeley, Berkeley, CA 94720, USA}
\address[nctu]{Institute of Physics, National Chiao-Tung University, Hsinchu, Taiwan}

\begin{abstract}
In this paper, we describe the design, construction and performance of an apparatus installed in the Aberdeen Tunnel laboratory in Hong Kong for studying spallation neutrons induced by cosmic-ray muons under a vertical rock overburden of 611 meter water equivalent (m.w.e.). The apparatus comprises of six horizontal layers of plastic-scintillator hodoscopes for determining the direction and position of the incident cosmic-ray muons. Sandwiched between the hodoscope planes is a neutron detector filled with 650 kg of liquid scintillator doped with about 0.06\% of Gadolinium by weight for improving the efficiency of detecting the spallation neutrons. Performance of the apparatus is also presented.
\end{abstract}

\end{frontmatter}


\section{Introduction}\label{sec:intro}
Neutrons are an important background for underground experiments studying neutrino oscillation, dark matter, neutrinoless double beta decay and the like. Majority of the neutron background is created by the ($\alpha$, $n$) interaction, where the $\alpha$ particles come from the decays of radioisotopes in the vicinity. 

Neutrons can also be created by interactions of cosmic-ray muons with matter in the underground laboratories. Contrary to neutrons coming from ($\alpha$, $n$), these spallation neutrons have a very broad energy distribution that extends to GeV. They can travel a long distance from the production vertices, and penetrate into the detector without being vetoed. As a result, they may be captured after thermalization, or interact with the detector materials to create fake signals. 

Since the spallation process is very complicated, it is highly desirable to investigate experimentally the production properties of the muon-induced spallation neutrons in underground environments. Such studies have been carried out in several underground experiments at different depths, ranging from 20 m.w.e. to 5,200 m.w.e. \cite{bib:Parlo_Verde}-\cite{bib:LSD}. The lack of experimental information also compromises the validity of simulation on spallation neutron.

The goal of the Aberdeen Tunnel experiment in Hong Kong is to study the production of spallation neutrons by cosmic-ray muons at a vertical depth of 235 m of rock (611 m.w.e.) using a tracking detector for tagging the incoming muons and a neutron detector filled with a liquid scintillator loaded with Gadolinium (Gd) for detecting the spallation neutrons.
In this paper, we present the details of the Aberdeen Tunnel experiment in Section 2. We will describe the design and construction features of the muon tracker and the neutron detector. In Section 3, calibration of the hodoscopes on the muon tracker and PMTs in the neutron detector will be discussed. The energy scale calibration of the neutron detector will be described. Data acquisition and trigger formation of the detectors will be presented. In Section 4, details of simulations and detector performance will be given.

\section{The Aberdeen Tunnel Laboratory}\label{sec:lab}
\subsection{Geological information}\label{sec:overview}
\begin{figure}[h]
\begin{center}
\includegraphics[scale=0.33]{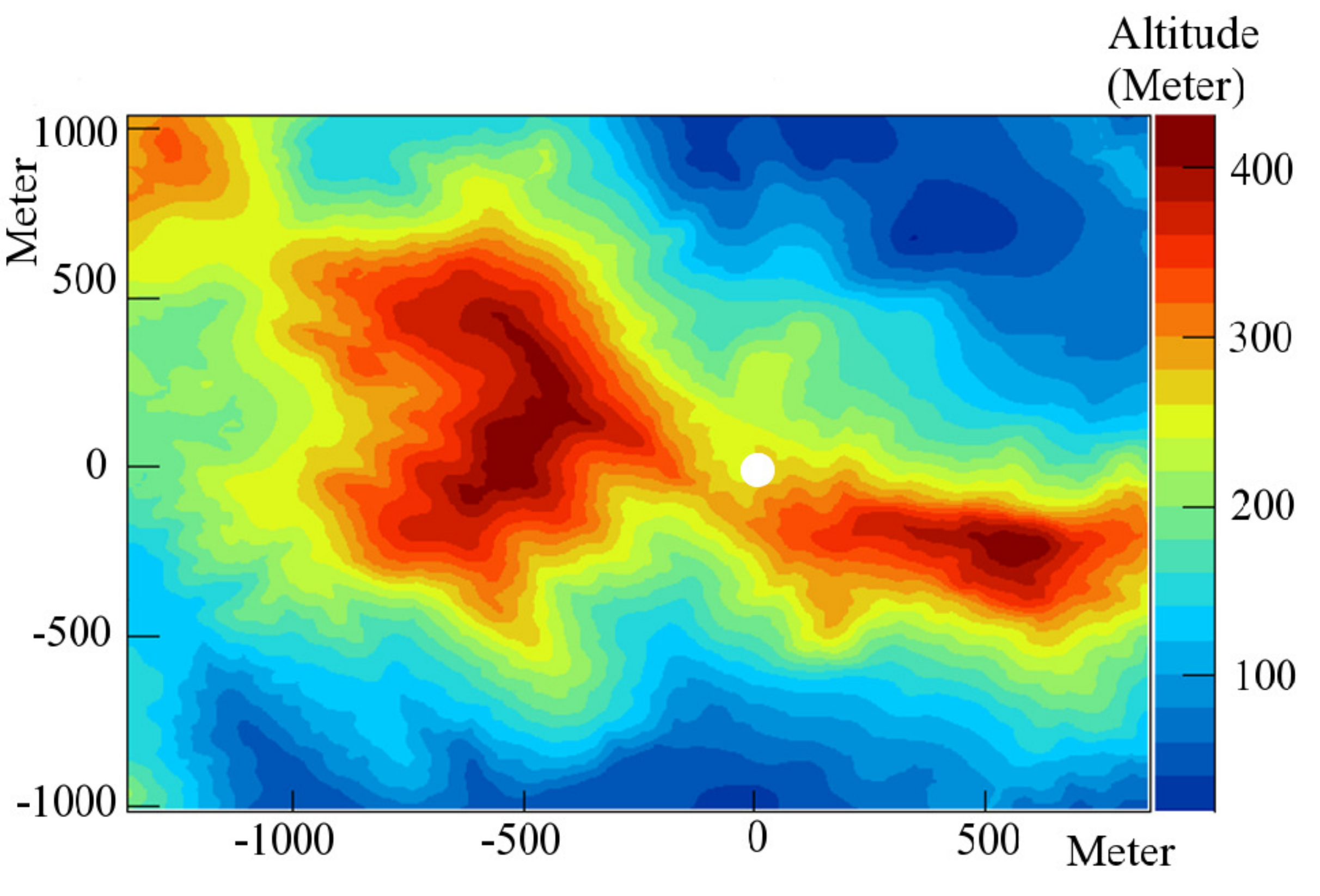}		
\caption{Contour map of the hills above the Aberdeen Tunnel underground laboratory denoted by a white dot at (0, 0). North is up. }
\label{fig:1_topo_map_2}
\end{center}
\end{figure}

The underground laboratory (at 22.23$^{\circ}$N and 114.6$^{\circ}$E) was constructed inside the Aberdeen Tunnel in Hong Kong Island in the early 1980s. The tunnel is a two-tube vehicle tunnel of 1.9-km long. It lies beneath the saddle-shape valley between two hills of over 400 m (Fig.~\ref{fig:1_topo_map_2}), namely Mount Cameron on the west and Mount Nicholson on the east.  The saddle-shape terrain provides a rock overburden of 611 m.w.e. for the laboratory which is located at the mid-point of the tunnel. It has dimensions of 6.7 m (L) $\times$ 3.2 m (W) $\times$ 2.2 m (H). The entrance is in a cross passage connecting the two traffic tubes. Access of the laboratory is only possible when all the traffic are diverted to one of the tubes during tunnel maintenance from 00:00 hr to 05:00 hr, typically two to four times a week.

\begin{figure}[h]
\begin{center}
\includegraphics[scale=0.32]{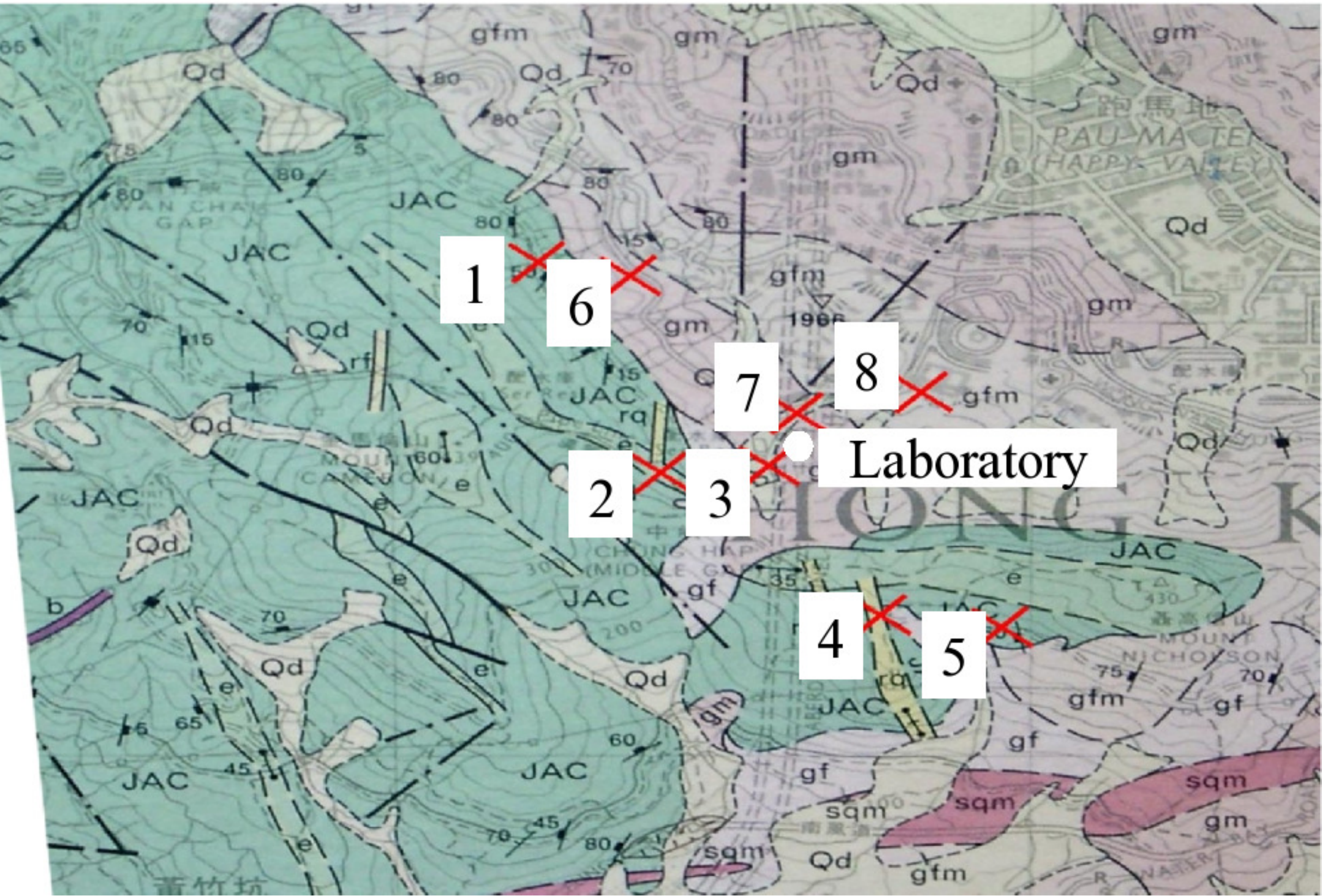} 
\caption{Geology in the vicinity of the Aberdeen Tunnel. Major types of rock are granite (gm, gfm), fine ash vitric tuff (JAC) and some debris flow deposits (Qd) \cite{bib:geology_map}. The dotted lines show the location of the Aberdeen Tunnel. }
\label{fig:2_topo_map_3}
\end{center}
\end{figure}

The major types of rock covering the Aberdeen Tunnel are granite and vitric tuff (Fig.~\ref{fig:2_topo_map_3}). In order to find out their chemical composition, rock samples were collected on the surface near the Aberdeen Tunnel for analysis. These rock samples were picked up at eight locations on the hiking trails in the area, as indicated in Fig.~\ref{fig:2_topo_map_3}. Samples of suitable size (from locations 2, 3, 7 and 8) were analyzed by X-ray fluorescence spectroscopy (XRF) at the University of Hong Kong. The chemical compositions are tabulated in Table \ref{table:rock_composition}. Determination of their physical properties was done at the Lawrence Berkeley National Laboratory \cite{bib:LBL_rock_test}. The results are shown in Table \ref{table:rock_phys}. Since these rock samples have been weathered, their physical properties may not reflect truly those of the rock inside the underground laboratory.

\begin{table}
\begin{center}
\begin{tabular}{c c c }
\hline
 Oxides & Composition (\%) & Error (\%) \\
\hline \hline
Silicon & 76.8 & 6.7 \\
Aluminum & 12.3 & 3.5 \\
Iron & 1.2 & 0.3 \\
Sodium & 1.4 & 1.3 \\
Potassium & 3.7 & 2.0 \\
\hline
\end{tabular}
\caption{Predominant composition of rock samples collected on the surface near the Aberdeen Tunnel.}
\label{table:rock_composition}
\end{center}
\end{table}

\begin{table*}
\begin{center}
\begin{tabular}{ c c c c c c c c}
\hline
Location 	& Bulk & Grain & P-wave & S-wave & Young's & Shear & Porosity   \\
		& density & density & velocity & velocity & modulus & modulus & (\%)  \\
		& (g cm$^{-3}$) & (g cm$^{-3}$)& (km/s) & (km/s)  & (GPa) & (GPa) & \\
\hline \hline
2	&	2.38	&	2.61 &	3.27		&	2.11		&	24.4	&	10.7	&	8.86		\\
3	&	2.57	&	2.62	&	5.56		&	3.29		&	68.5	&	27.8	&	1.86		\\
7	&	2.36	&	2.61 &	3.24		&	2.00		&	22.7	&	9.50	&	9.55		\\
8	&	2.47	&	2.59 &	4.28		&	2.68		&	41.9	&	17.8	&	4.58		\\
\hline
\end{tabular}
\caption{Physical properties of rock samples collected on the surface at various locations in Fig.~\ref{fig:2_topo_map_3} near the Aberdeen Tunnel \cite{bib:LBL_rock_test}. }
\label{table:rock_phys}
\end{center}
\end{table*}

Using a modified Gaisser's parametrization \cite{bib:guanmengyuen} for generating the energy of the cosmic-ray muons on the surface, a digitized three-dimensional topographical map with a resolution of 10 m and area of 13.6 km by 10.2 km, and MUSIC \cite{bib:music} for propagating the muons to the location of the underground laboratory, the mean energy of the muons getting to the laboratory is estimated to be 120 GeV and the integrated flux is approximately $9.6 \times 10^{-6}$ cm$^{-2}$ s$^{-1}$.

\subsection{Laboratory environment}\label{sec:lab_env}

Humidity and temperature of the laboratory are monitored by sensors and the data can be accessed remotely through the internet. Ambient temperature in the laboratory is kept between 20$^{\circ}$C and 30$^{\circ}$C with a typical value of about 23$^{\circ}$C, relative humidity between 35\% and 40\% throughout the year. A surveillance camera is installed for monitoring the environment of the underground laboratory remotely.

The walls and floor are lined with cement and protective paint to reduce dust and radon emanation. The mean radon concentration in the laboratory\footnote{Measured with a high-sensitivity radon detector developed by the University of Hong Kong.} is 299 $\pm$ 20 Bq m$^{-3}$, with a range of 250 to 325 Bq m$^{-3}$. The neutron ambient dose equivalent \cite{bib:icrp60} is measured to be 0.7 $\pm$ 0.04 nSv/h with a Helium-3 detector in a polyethylene sphere of 25-cm diameter manufactured by Berthold Technologies GmbH \& Co. KG  (LB6411) \cite{bib:Berthold}.

\begin{figure}[h]
\begin{center}
\includegraphics[scale=0.4]{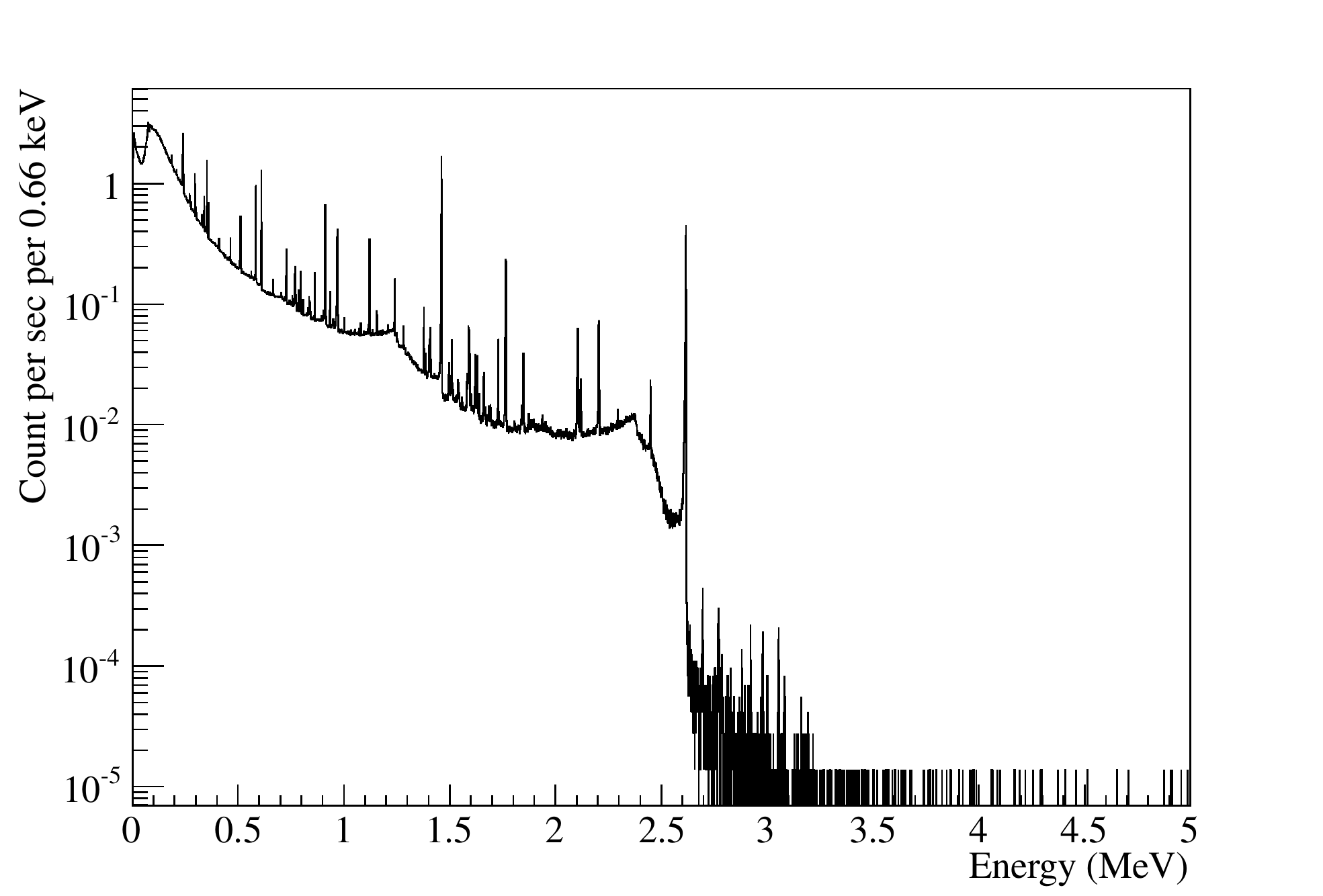}		
\caption{Energy spectrum of the gamma-ray background measured with a high-purity germanium detector inside the Aberdeen Tunnel laboratory.}
\label{fig:3_gamma_spectrum}
\end{center}
\end{figure}

Gamma-rays from primordial Uranium (U), Thorium (Th) and Potassium-40 ($^{40}$K) in the surrounding rock are the major sources of background in the Aberdeen Tunnel laboratory. The amount of ambient gamma-rays was measured \textit{in situ} for 20 hours with an Ortec GEM 35S high-purity germanium detector (HPGe) that has a cylindrical Ge crystal of 61.5 mm in diameter and 65.9 mm in height. The measured energy spectrum is shown in Fig.~\ref{fig:3_gamma_spectrum}.

Activities of $^{40}$K and radioisotopes of U and Th series in rock were calculated by simulation. Gamma-rays were generated uniformly from the surrounding rock, then propagated to the Ge crystal. Attenuation of the gamma-rays was calculated from the mass attenuation coefficient of the standard rock. By normalizing the simulation spectrum to match the experimental results, the simulated gamma-ray events were converted to the corresponding activities of the isotopes. At the surface of the rock, the flux of gamma-rays with energy up to 3 MeV was estimated to be 29~$\pm$~1~cm$^{-2}$~s$^{-1}$. With the approximation of secular equilibrium, activities of $^{238}$U and $^{232}$Th in the rock samples were calculated using the gamma-rays and their branching ratios in the same series. For $^{40}$K, its activity was deduced from the intensity of the 1.46-MeV gamma-ray line in the spectrum shown in Fig. 3. The results are summarized in Table~3.

\begin{table}[h]
\begin{center}
\begin{tabular}{ c c }
\hline
 Isotope & Activity (Bq/kg)  \\
\hline \hline
$^{238}$U & 85$\pm$2  \\
$^{232}$Th & 108$\pm$3  \\
$^{40}$K & 1007$\pm$60  \\
\hline
\end{tabular}
\caption{Activity of $^{238}$U, $^{232}$Th and $^{40}$K in the rocks surrounding the Aberdeen Tunnel laboratory.}
\label{table:gamma_conc}
\end{center}
\end{table}

\section{Apparatus}\label{sec:app}
The incoming cosmic-ray muons and the spallation neutrons are measured with two different detectors. A muon tracker (MT) determines the angular distribution and flux of the muons, while the neutron detector (ND) observes the spallation neutrons. Augmented with custom-built and commercially available electronics modules, a MIDAS-based data acquisition system 
is used to collect data.  Details of these subsystems are presented in this section.

\subsection{Muon tracker}\label{sec:muon}
\begin{figure*}
\begin{center} 
\includegraphics[scale=0.4]{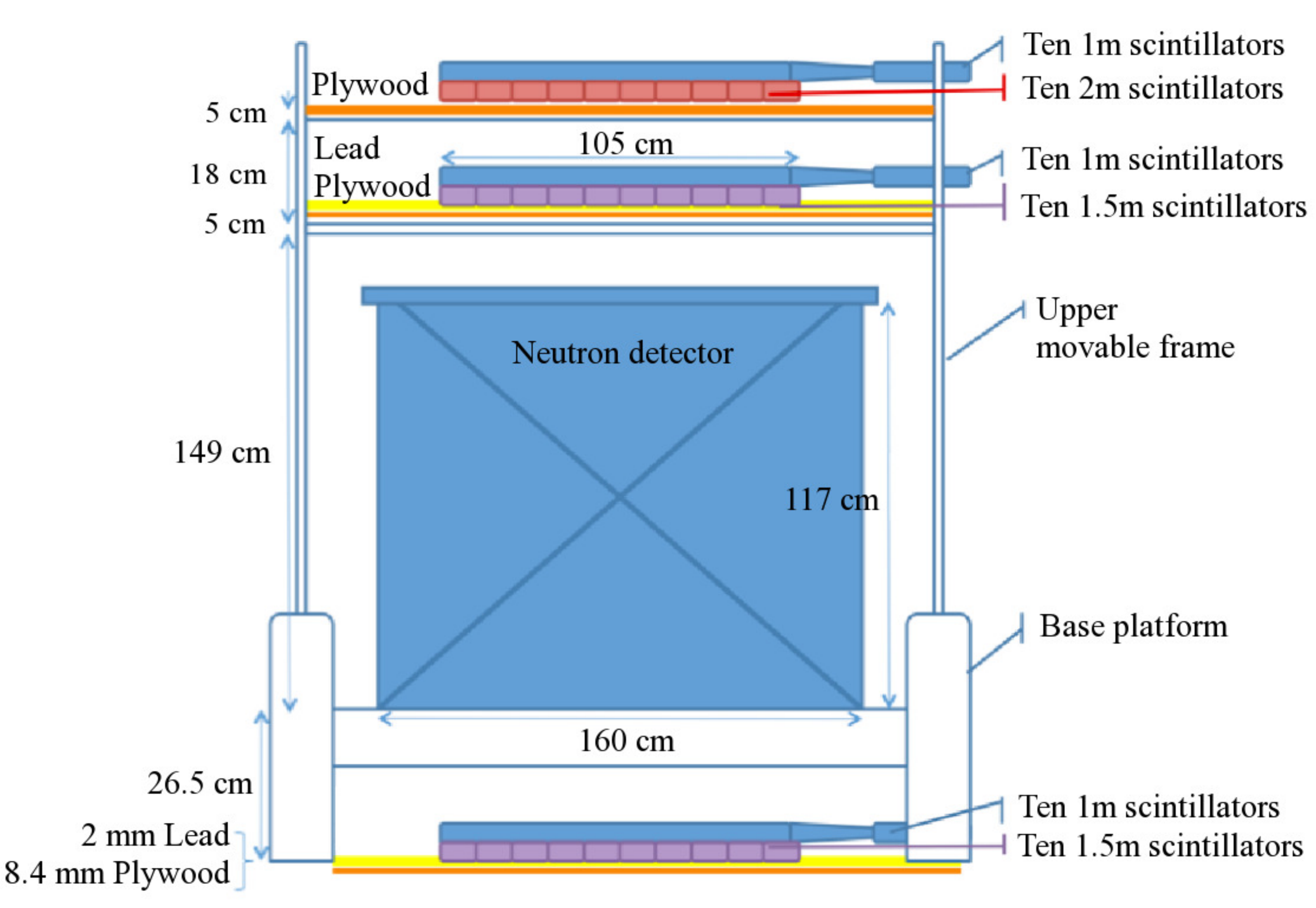} 
\caption{Front view of the MT and the ND. Right-hand side is east.}
\label{fig:4_mt_1}
\end{center}
\end{figure*}

\begin{figure*}
\begin{center}
\includegraphics[scale=0.4]{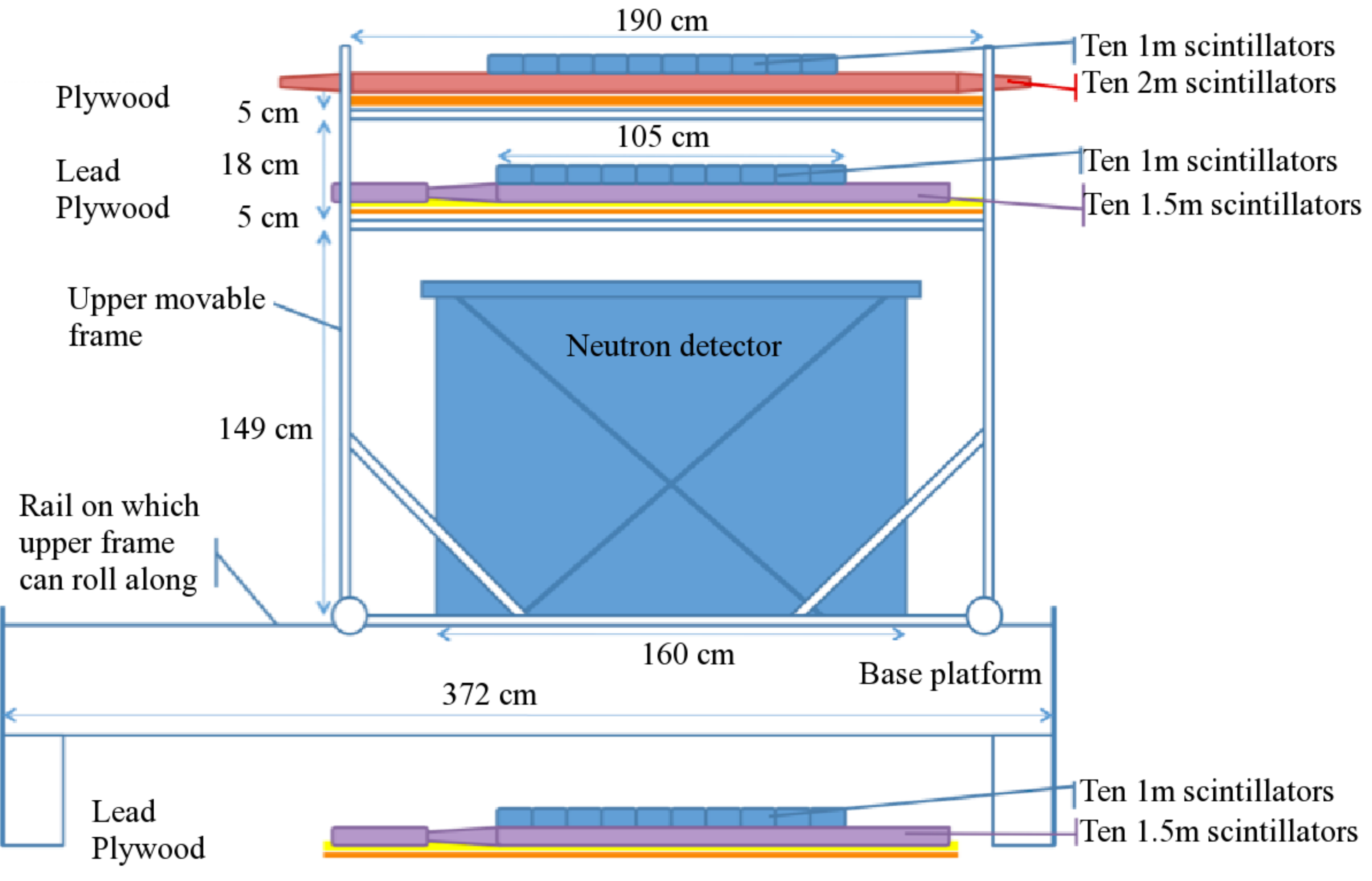} 
\caption{Side view of the MT and the ND. Right-hand side is north.}
\label{fig:5_mt_2}
\end{center}
\end{figure*}

The muon tracker (MT) consists of 60 plastic scintillator hodoscopes arranged in three layers as shown in Figs.~\ref{fig:4_mt_1} and \ref{fig:5_mt_2}. The separation between the top and bottom layer is 198 cm. Each layer is made up of two planes of hodoscopes orthogonal to each other for determining the (x,y) coordinates of a muon passage through the layer. 

\subsubsection{Muon tracker frame}
The plastic scintillator hodoscopes are supported by a steel frame. This frame consists of an upper support structure and a base platform. The upper structure can be slid on two parallel rails, each of which is 372-cm-long running in the north-south direction. The slidable structure supports two layers of hodoscope above the ND. When the upper frame is in the south-most position, the entire ND on the base platform is sandwiched by the three hodoscope layers. The configuration of the MT can be changed for measuring muons at larger zenith angles. This also facilitates the installation and calibration of the ND. 

\subsubsection{Top hodoscope layer} 
The top layer is formed by ten 1-m-long hodoscopes and ten 
2-m-long hodoscopes, as illustrated in Figs.~\ref{fig:4_mt_1} and \ref{fig:5_mt_2}. For the 1-m-long hodoscopes, each plastic scintillator has dimensions of 100 cm (L) $\times$ 10 cm (W) $\times$ 2.54 cm (T). A 5-cm-diameter PMT (Amperex XP2230) is attached to a Lucite light guide at one end. The whole array forms a sensitive area of (100 $\times$ 100) cm$^{2}$. Underneath the 1-m-long hodoscope layer lies ten 2-m-long hodoscopes, forming an active area of (93 $\times$ 200) cm$^{2}$. Each hodoscope is made of a piece of 200 cm (L) $\times$ 9.3 cm (W) $\times$ 2.54 cm (T) plastic scintillator. Two Hamamatsu H7826 photomultiplier tubes (SPMTs), each with a circular photocathode of 1.9 cm in diameter, are coupled to the two ends of each plastic scintillator with tapered Lucite light guides.

\subsubsection{Middle and bottom hodoscope layers} 
The middle and bottom layers have identical configuration. In each layer, ten 1-m-long plastic scintillators lie in the east-west direction, each has a 5-cm Amperex XP2230 PMTs at the eastern end. The whole plane has a sensitive area of (100 $\times$ 100) cm$^{2}$. Underneath this hodoscope plane are ten 1.5-m-long hodoscopes, lying orthogonally to the 1-m-long ones, with Amperex XP2230 PMTs in the north. Dimensions of the 1.5-m-long scintillator are 150 cm (L) $\times$ 10 cm (W) $\times$ 2.54 cm (T). They form a sensitive area of (100 $\times$ 150) cm$^{2}$.

\subsection{Neutron detector}\label{sec:neutrondetector}
\begin{figure}[h]
\begin{center}
\includegraphics[scale=0.3]{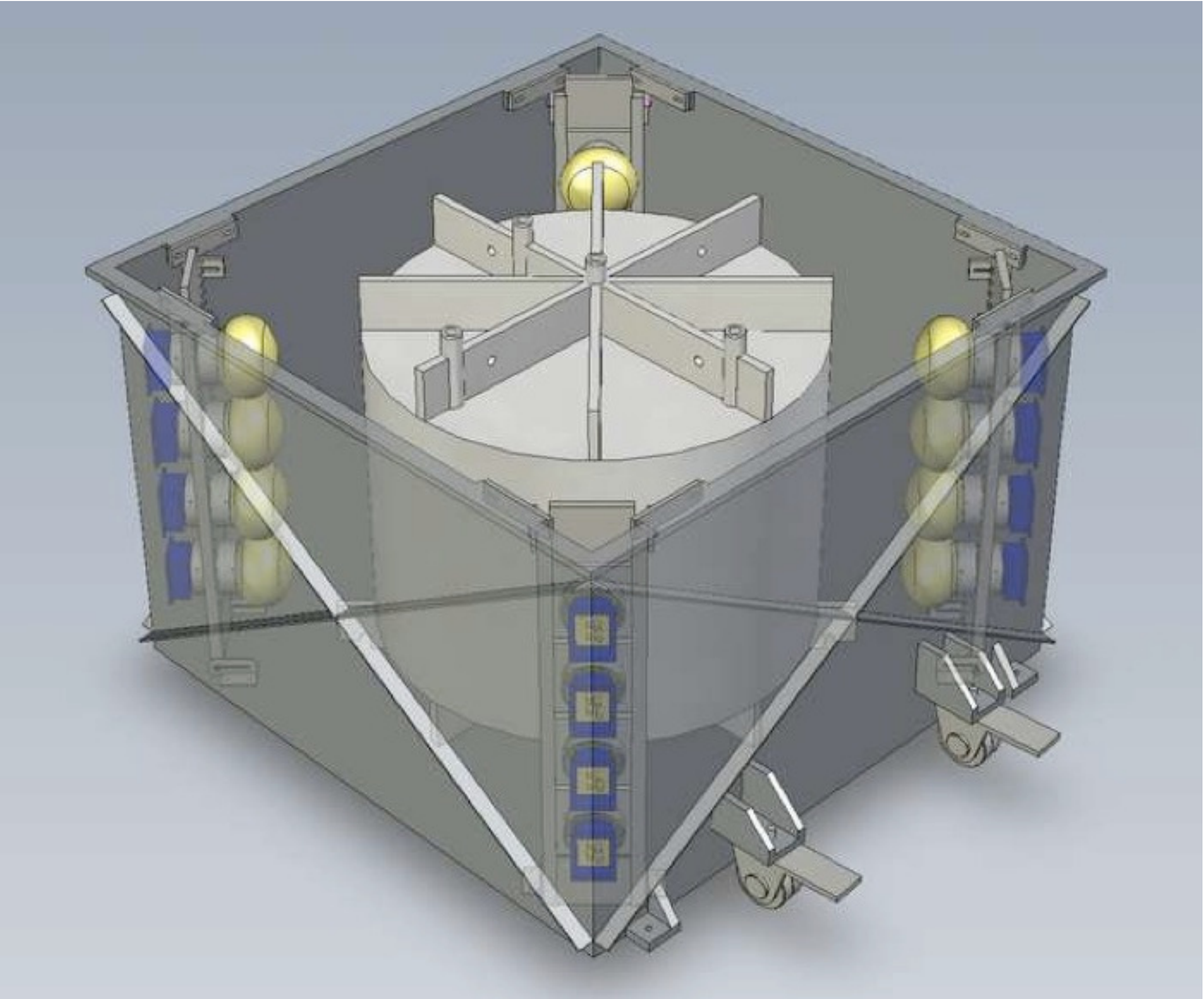}
\caption{Schematic drawing of the ND. The cylindrical acrylic vessel and the stainless steel rectangular tank are shown. Sixteen 20-cm PMTs are located at the four corners of the stainless steel tank. Top and bottom reflectors are not shown.}
\label{fig:6_nd}
\end{center}
\end{figure}

In the space sandwiched by the hodoscope layers, a neutron detector (ND) is set up for detecting muon-induced neutrons. The detector employs a 2-zone design (Fig.~\ref{fig:6_nd}). In the inner zone, an acrylic vessel is filled with 760 L (650~kg) of liquid scintillator as the target. The liquid scintillator is loaded with 0.06\% of gadolinium (Gd) to enhance neutron-capture. The acrylic vessel full of Gd-doped liquid scintillator (Gd-LS) is submerged in 1,900 L (1,630~kg) of mineral oil that serves as the outer shield for suppressing the amount of ambient gamma-rays and thermal neutrons entering the Gd-LS. The mineral oil and Gd-LS are sealed in a rectangular stainless steel tank, keeping them in a light-tight condition and free from oxygen in the atmosphere. 

When gadolinium in Gd-LS captures a neutron, it produces a gamma cascade with a total energy of about 8 MeV. Scintillation photons created by the gamma-rays are collected with sixteen Hamamatsu R1408 20-cm PMTs in the ND. Reflectors are installed at the top and at the bottom of the acrylic vessel for improving the light-collection efficiency and to have a more uniform energy response. 

\subsubsection{Acrylic vessel}\label{subsubsec:av}
The Gd-LS is held in a cylindrical acrylic vessel manufactured by Nakano International Co., Ltd. in Taiwan \cite{bib:nakano}. UV-transmitting acrylic is chosen for its compatibility with Gd-LS and higher transmittance in the UV-visible region. For the curved surface of the vessel, which is 1-cm-thick, the optical transmittance is above 86\% for wavelength longer than 350 nm. This is crucial for achieving high efficiency for detecting scintillation photons with the Hamamatsu R1408 PMTs that have good quantum efficiency between 350 and 480 nm.

\begin{figure*}[ht]
\begin{center}
\includegraphics[scale=0.35]{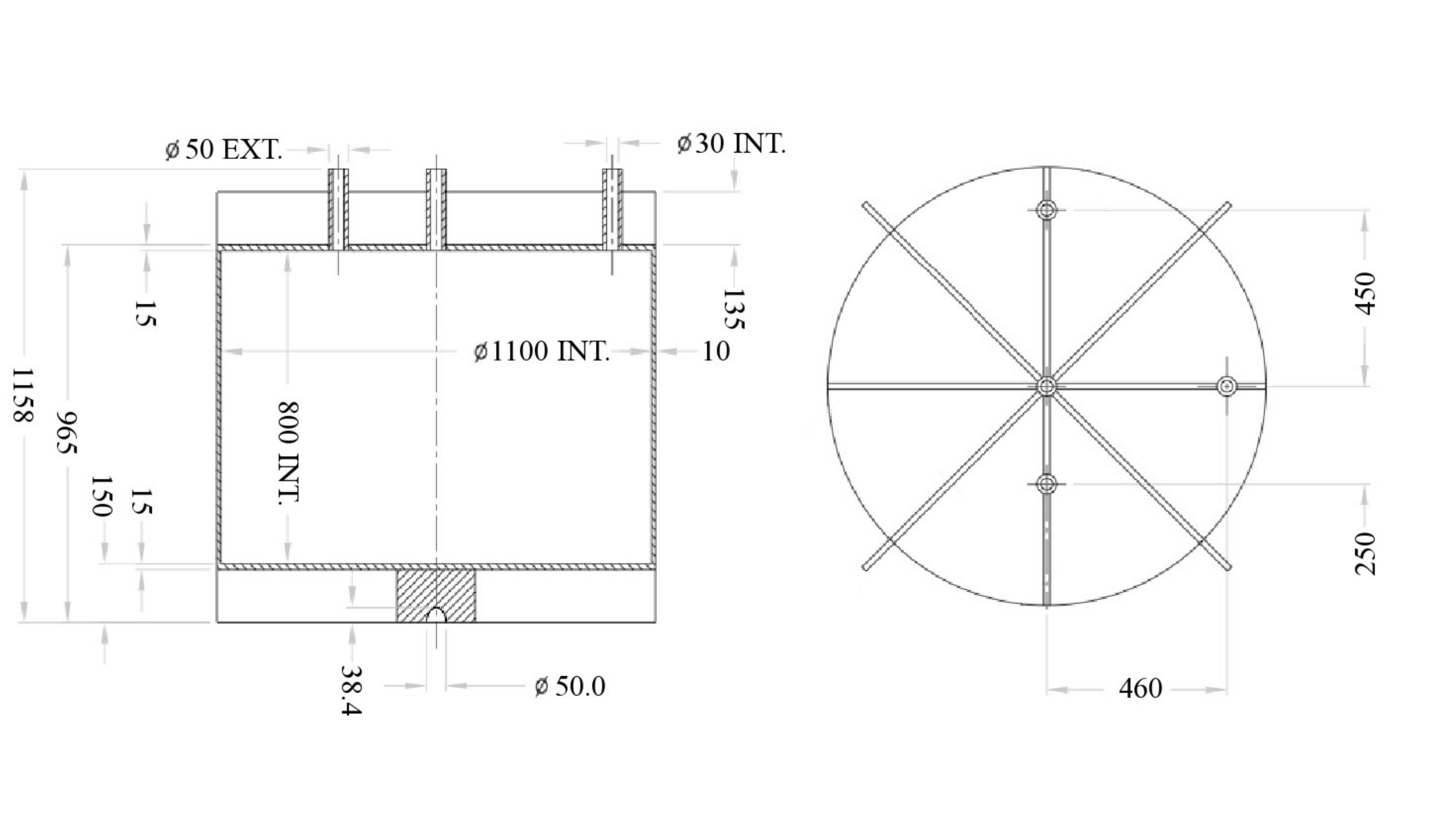}
\caption{Acrylic vessel in the ND (unit in millimeters). Left: Side view of the vessel along the calibration ports. Right: Top view showing the locations of three calibration ports and one overflow port (46 cm from center).}
\label{fig:7_av}
\end{center}
\end{figure*}

On the top surface of the acrylic vessel, three calibration ports are opened for the deployment of calibration sources (Fig.~\ref{fig:7_av}). The calibration ports are located at different radial distances from the center (0 cm, 25 cm and 45 cm). There is also an overflow port leading to an overflow tank for accommodating the thermal expansion of the Gd-LS in the vessel.

Both the top and bottom plates of the acrylic vessel are 1.5 cm thick. 
As shown in Fig.~\ref{fig:7_av}, each plate has eight 1-cm-wide ribs, extending radially for structural reinforcement. The bottom ribs raise the Gd-LS such that mineral oil can fill the 13.5-cm space below the acrylic vessel. The top ribs provide support to the reflector above the acrylic vessel.

\subsubsection{Stainless steel tank}\label{subsubsec:ss_tank}
The stainless steel tank is responsible for holding all elements of the ND intact and keeping them in a light- and air-tight environment. The inner dimensions of the tank are 160 cm (L) $\times$ 160 cm (W) $\times$ 117.3 cm (H). The interior of the ND is painted with a black fluoropolymer paint which is compatible with the mineral oil.

On the top lid of the stainless steel tank, three flanges are installed for the calibration ports of the acrylic vessel. There are gate valves on the flanges such that the port can be opened for deployment of calibration sources.

At the four corners of the stainless steel top lid, there are patch panels for deploying the 20-cm PMTs. The patch panels have hermetic feedthroughs for connecting the signal cables and high-voltage cables of the ND PMTs. 

The relative position of the acrylic vessel and the stainless steel tank is fixed by an anchor welded to the bottom of the stainless steel tank. This hemispherical anchor has a size and shape matching the hollow space at the center of the bottom ribs of the acrylic vessel. Once the acrylic vessel is placed in the stainless steel tank, it couples to the anchor. Only a small rotation of the acrylic vessel is possible afterwards.

\subsection{Gadolinium-doped liquid scintillator}
The Gd-LS based on the recipe described in \cite{bib:yeh} was synthesized in Hong Kong. To increase the scintillation efficiency and to shift the emission spectrum to the sensitive region of the Hamamatsu R1408 PMTs, 1.3~g/L of 2,5~diphenyloxazole (PPO) and 6.7 mg/L of $p$-bis-($o$-methylstyryl)-benzene (bis-MSB) were added to the Gd-LS as primary and secondary fluors. The major solvent is linear alkylbenzene (LAB), which is sold under the commercial name of Petrelab 550-Q, by Petresa, Canada \cite{bib:petresa}. LAB is chemically less active and has a higher flash point than the other liquid scintillators like pseudocumene, while the emission spectrum and light yield are comparable \cite{bib:dybtdr}.

The molecule of Petrelab 550-Q has a benzene ring attached to an alkyl derivative that contains 10 to 13 carbon atoms. The carbon and hydrogen atoms can also capture thermal neutrons with cross-sections of 0.00337 barns and 0.332 barns respectively \cite{bib:yeh}. The capture time of neutron on hydrogen is about 200 $\mu$s. The energy of the gamma-ray emitted in the subsequent nuclear de-excitation is 4.95 MeV for carbon and 2.22 MeV for hydrogen. Doping Gd in liquid scintillator enhances neutron capture significantly because two of the isotopes of Gd have much larger thermal neutron-capture cross-section of 60,900 barns ($^{155}$Gd) and 254,000 barns ($^{157}$Gd) at 0.0253 eV \cite{bib:mughabghab}. With 0.06\% of Gd, by weight, added to the liquid scintillator, the neutron capture time is shortened to about 50 $\mu$s. Moreover, the total energy of the emitted gamma-rays  is about 8 MeV. This provides a powerful criterion for discriminating the neutron-capture signals against the ambient gamma-rays that have energies below 2.6 MeV.

\subsubsection{Photomultiplier tubes}\label{subsubsec:pmt_cal}
\begin{figure}[h]
\begin{center}
\includegraphics[scale=0.25]{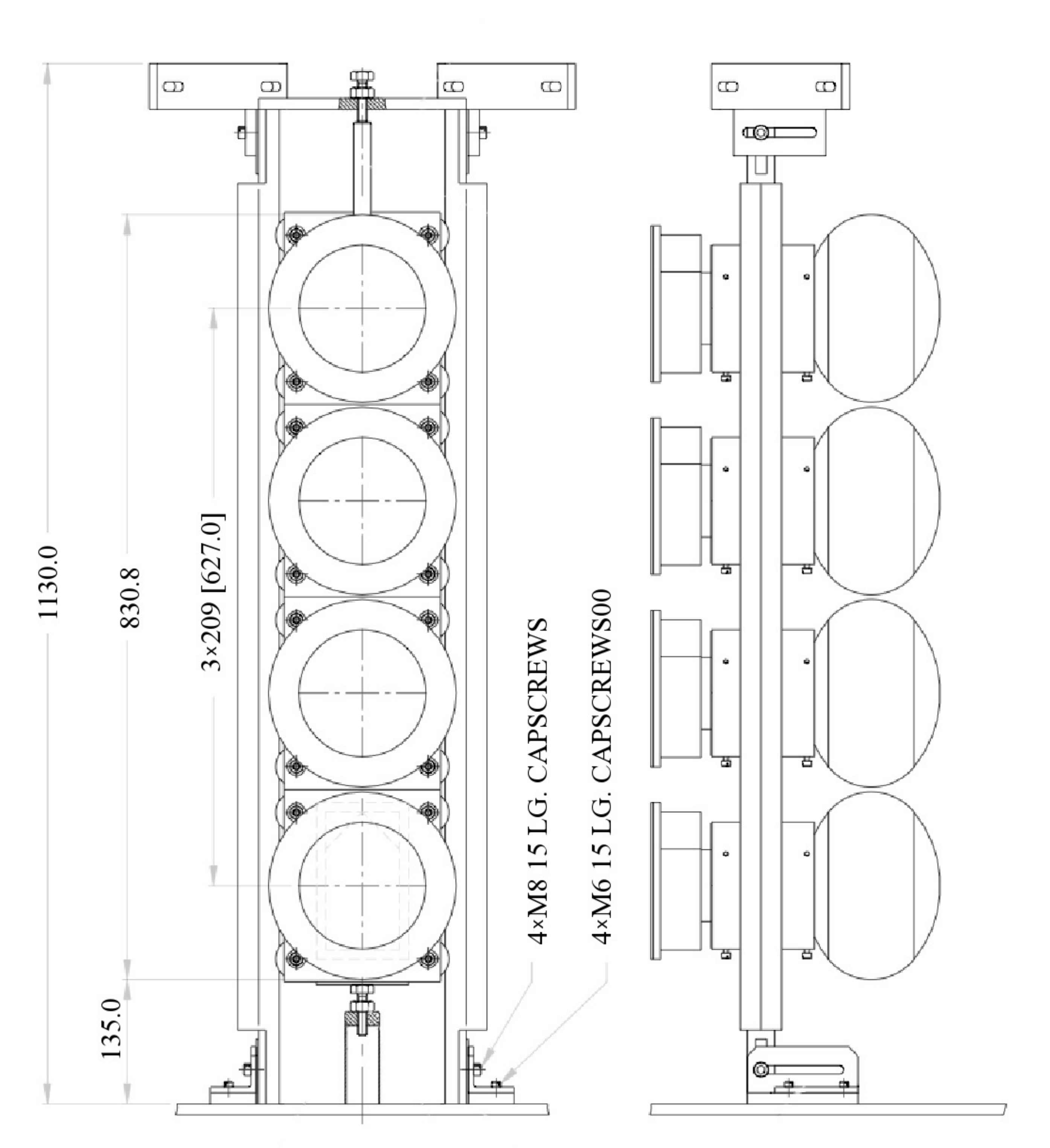}
\caption{Drawings of PMT rails and PMT mounts inside the stainless steel tank (unit in millimeters). }
\label{fig:8_pmt_rail}
\end{center}
\end{figure}

Sixteen Hamamatsu R1408 20-cm PMTs are used in the ND.  
R1408 is the predecessor of the newer model R5912. The R1408 PMT has a hemispherical photocathode. The high-voltage divider in the PMT base was designed and made by the MACRO collaboration. Stainless steel PMT mounts are used to hold the PMTs in place and seal the PMT base from the mineral oil. In addition, the PMT mounts have wheels for deploying the PMTs into the mineral oil by sliding along the vertical rails below the patch panels of the stainless steel tank (Fig.~\ref{fig:8_pmt_rail}). The alignment of the PMT rails was adjusted with a laser pointer. To reduce the amount of light scattering, the PMT mounts and rails are all coated with black fluoropolymer paint.  The PMTs were installed in four columns at the corners of the stainless steel tank. All the PMTs are equally spaced, and point to the vertical central axis of the acrylic vessel at the same radial distance.

\subsubsection{Reflectors}\label{subsubsec:reflectors}
\begin{figure}[h]
\begin{center}
\includegraphics[scale=0.22]{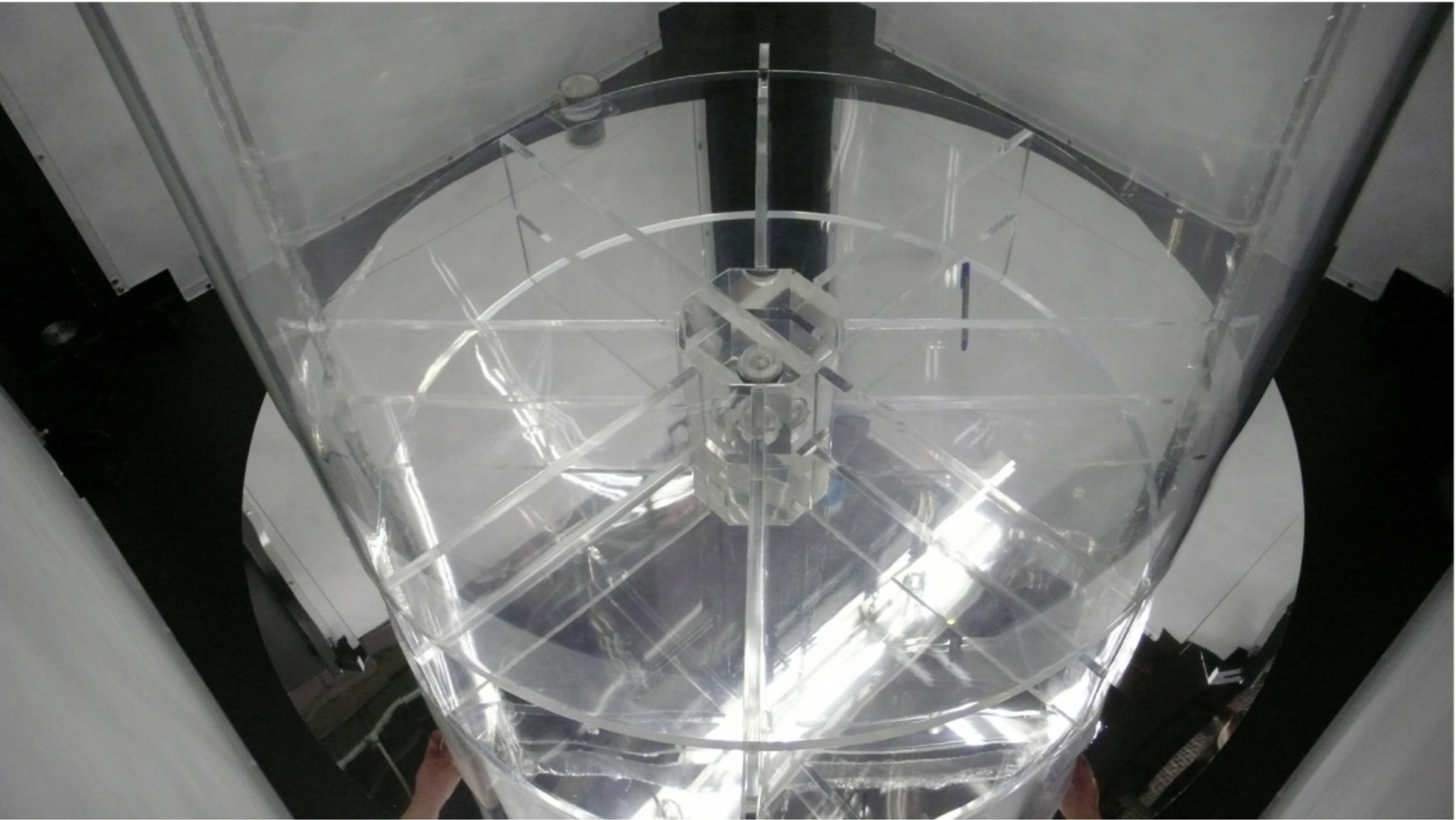}
\caption{Reflectors inside the neutron detector. Tyvek sheets are mounted on the walls in the stainless steel tank. The specular bottom reflector is under the acrylic vessel. The top specular reflector is not shown here.}
\label{fig:9_tyvek}
\end{center}
\end{figure}

Four panels of white diffuse reflectors are mounted to the inner walls of the stainless steel tank. They are made of DuPont Tyvek 1085D mounted on 8-mm-thick acrylic plates. DuPont Tyvek 1085D is selected for its appropriate thickness, reflectivity and availability \cite{bib:tyvek_bnl}. The four corners of the Tyvek panels are hung on the PMT rails (Fig.~\ref{fig:9_tyvek}).
Two circular reflectors of 140-cm-diameter are put on the top and at the bottom of the ribs of the acrylic vessel. They increase light collection of the PMTs by specular reflection. The size of the reflectors is maximized such that they would not hinder the deployment of the PMTs through the patch panels. 
The circular reflectors were made by gluing reflective Miro-Silver pieces \cite{bib:miro_silver}  to an acrylic backing plate with DP810 glue from 3M. The Miro-Silver is a 0.2-mm-thick aluminum sheet coated with super reflective oxide-layer which has a reflectivity of over 96\% $\pm$ 1\% in air and 95\% $\pm$ 1\% in mineral oil at a wavelength of 532 nm.

\begin{figure*}[t]
\begin{center}
\includegraphics[scale=0.3]{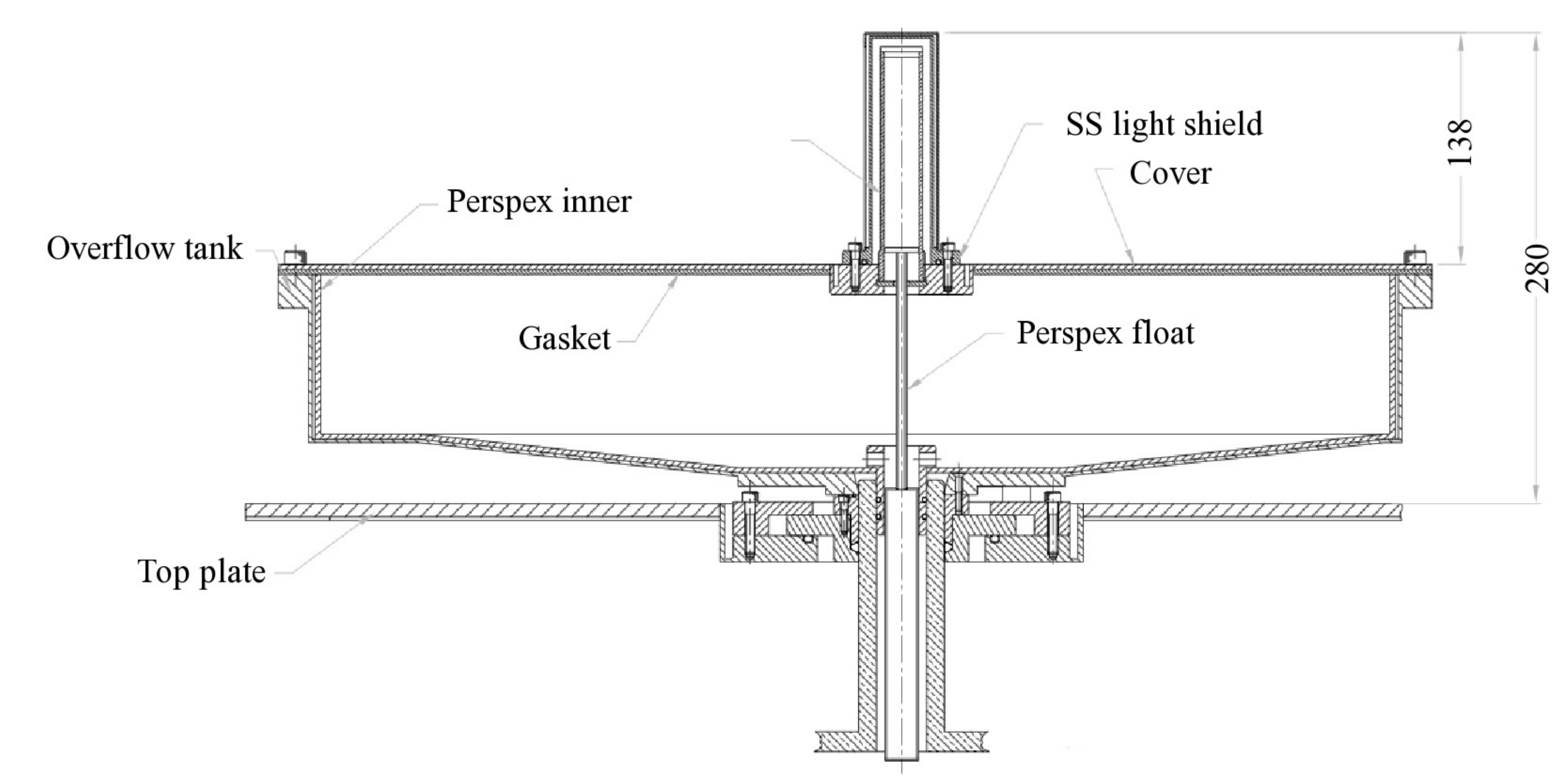}
\caption{Gd-LS overflow tank of the ND (unit in millimeters).}
\label{fig:10_overflow}
\end{center}
\end{figure*} 
\subsubsection{The overflow tank}\label{subsubsec:overflow}
A 15-L overflow tank is connected to the acrylic vessel (Fig.~\ref{fig:10_overflow}). As the thermal expansion of Gd-LS in the ND equals 0.59 L K$^{-1}$ \cite{bib:dybtdr}, this reservoir is able to hold the extra volume of Gd-LS for a temperature rise of 25$^{\circ}$C.

The overflow tank is made of stainless steel lined with an inner acrylic layer. Air inside the tank is displaced with a slow flow of nitrogen gas. Liquid level in the overflow tank is revealed by a hollow acrylic indicator floating in the Gd-LS inside the overflow tube. An observing window can be opened to check the position of the indicator. An EVOH bag is connected to the overflow tank as a pressure relief for the expanding Gd-LS and nitrogen gas when the temperature rises.

To take care of the thermal expansion of the mineral oil, a gap below the stainless steel top lid filled with nitrogen gas serves as a buffer. The gas gap is 23-mm-tall, yielding a volume of more than 50 L. This can hold the additional volume of mineral oil for a temperature rise of 25$^{\circ}$C.

\subsection{Calibration system}\label{sec:nd_cal}
A deployment box is used to place calibration sources at different positions through the three ports in the ND. The box has a flange on its bottom surface, which fits to the gate valves of the calibration ports. Inside the box, the source holder and all the wetted parts are made of acrylic or covered by materials compatible with Gd-LS to prevent etching. 

\begin{figure}[h]
\begin{center}
\includegraphics[scale=0.35]{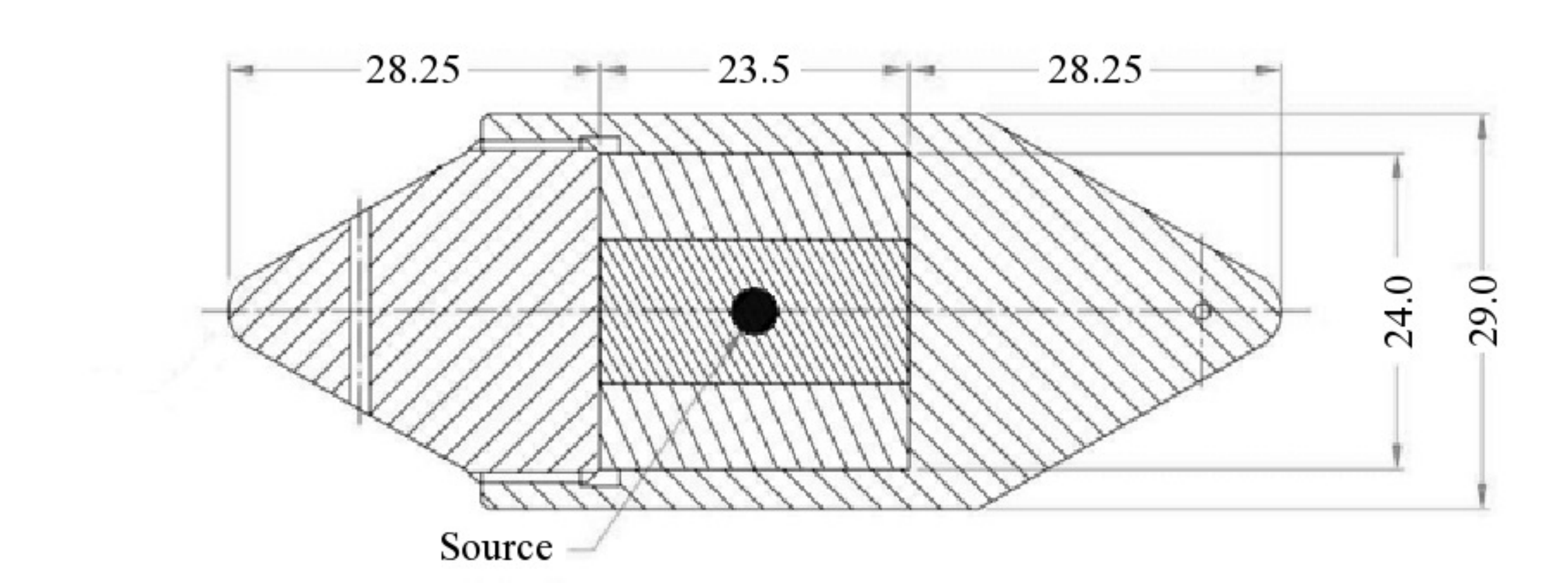}
\caption{Acrylic holder for deploying radioactive calibration sources into Gd-LS (unit in millimeters).}
\label{fig:11_src_holder}
\end{center}
\end{figure}

Each of the radioactive source holder and LED light sources are attached to a thin wire wound around an acrylic winch. The wire is protected by fluoropolymer heat-shrink tubing. The winch is engaged to a stepping motor that controls the vertical position of the calibration source. A separate winch is made for the radioactive sources and each of the light sources. Sources can be swapped easily by replacing the whole winch assembly.
An acrylic holder (Fig.~\ref{fig:11_src_holder}) is used to seal and to keep the radioactive source away from the Gd-LS. The screw-cap can be opened to replace the radioactive source inside.

\begin{figure}[h]
\begin{center}
\includegraphics[scale=0.25]{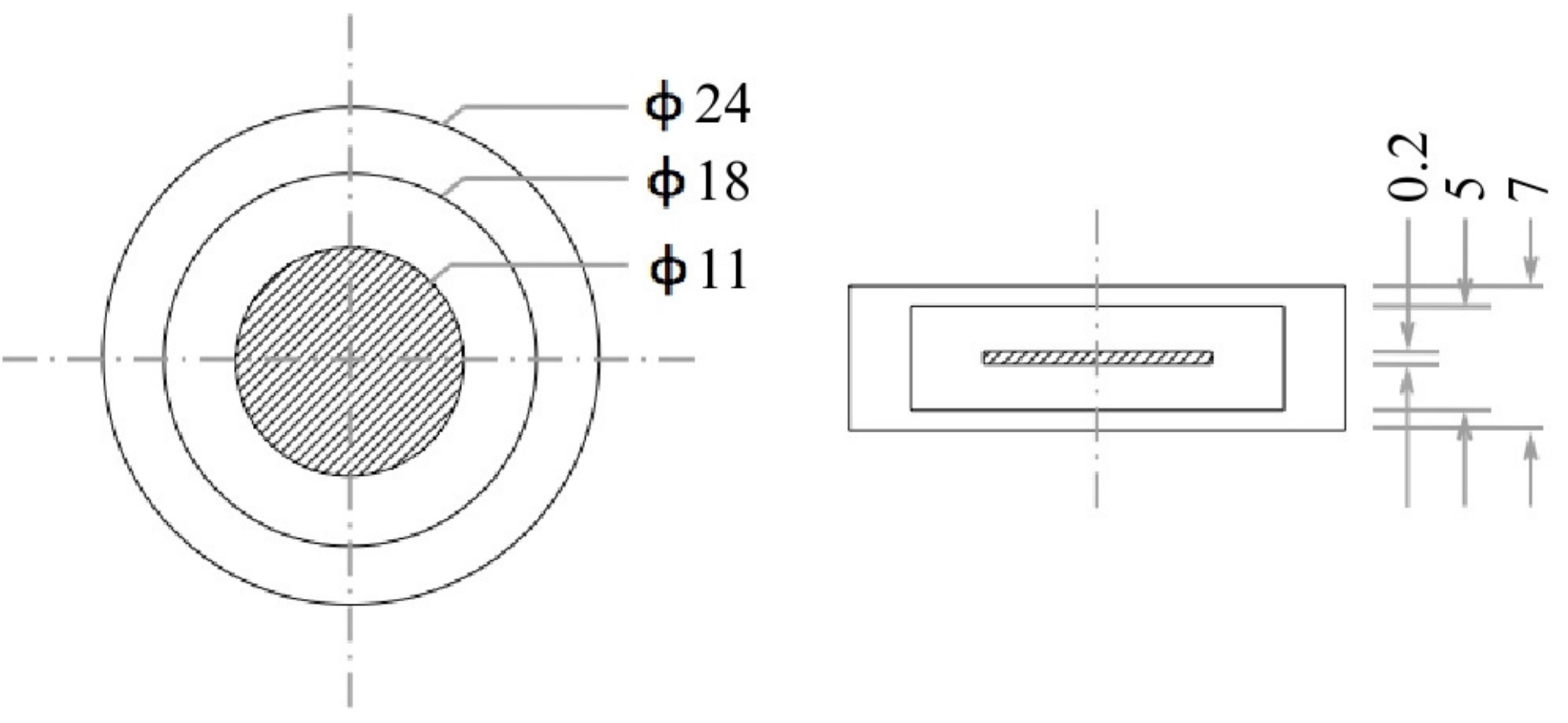}
\caption{Geometry of the Am-Be source (unit in millimeters). Shaded region is the active region, encapsulated in a plastic shell, and an outer Molybdenum shell.}
\label{fig:12_AmBeGeom}
\end{center}
\end{figure}

Radioactive sources are used for calibrating the energy scale of the ND. Cobalt-60 ($^{60}$Co) providing 1.17-MeV and 1.33-MeV gamma-rays, Cesium-137 ($^{137}$Cs) emitting 0.66-MeV gamma-rays, and Americium-Beryllium (Am-Be) neutron source are used. The construction of the Am-Be source is shown in Fig.~\ref{fig:12_AmBeGeom}. Prominent ambient gamma-rays (1.46 MeV $\gamma$ from Potassium-40, and 2.61 MeV $\gamma$ from Thallium-208) are also used for real-time calibration. 

LED sources are used for checking the linear response of the ND. Two kinds of light sources are available: an isotropic source and a planar source. For the isotropic source, a Teflon ball is used as a diffuser for the two LEDs embedded inside. The calibration system sends triggers at different time to the Sheffield pulser \cite{bib:sheffield_pulser} of each LED to control the flashing sequence. The bias voltage of the LED is also adjustable for producing different light intensities. The planar source employs the same electronics as the isotropic source. The Teflon diffuser of this source is partly hidden behind a horizontal slit of adjustable size such that a horizontal plane of light is generated. It is used to cross-check the relative gain of the PMTs in the same horizontal plane. 

The deployment box is connected to a computer. Motion of the stepping motor, the bias voltage and pulsing frequency of each LED are controlled through the computer. An infrared camera and infrared LEDs are also installed in the light-tight deployment box to view the passage of the sources through the narrow calibration port. All infrared LEDs are switched off during data taking.

The deployment box is sealed with o-rings and draw-latches to prevent oxygen in air from entering the ND. Moreover, a purging system is connected to it. Nitrogen is used to purge the box after it is opened for changing the source. After purging with nitrogen of three
volumes of the box, the gate valve is then opened for deploying the source.

\subsection{Data acquisition system}\label{sec:daq}
\begin{figure*}
\centering
\includegraphics[scale=0.5]{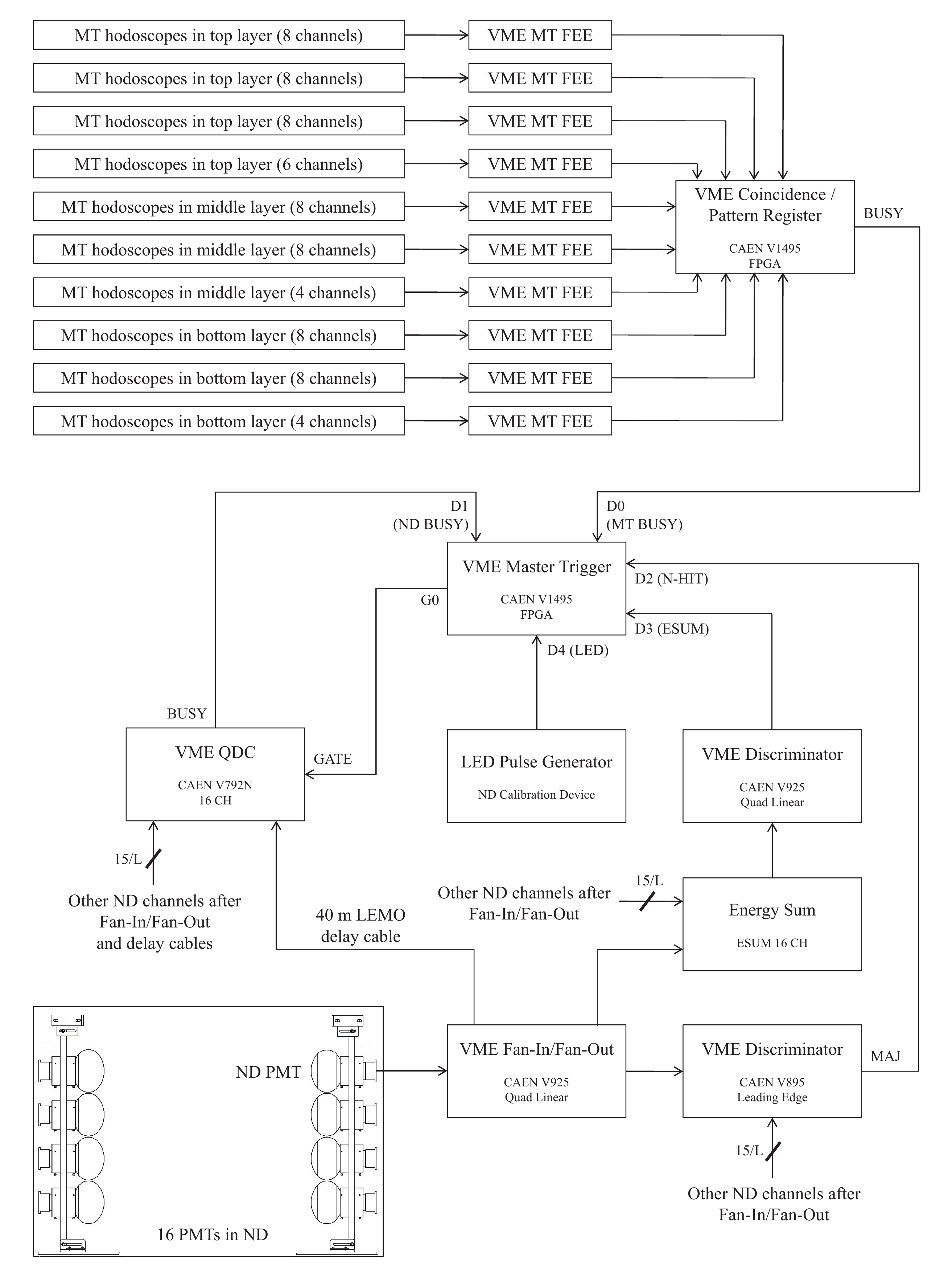}
\caption{Block diagram of the data acquisition system.}
\label{fig:13_daq-block-diagram}
\end{figure*}

The DAQ setup is shown in Fig.~\ref{fig:13_daq-block-diagram}. The Front-End Electronics (FEE) is used to process the PMT output signals. The essential functions of FEE are as follows: 
\begin{itemize}
\item Provide fast information to the trigger system
\item Provide precision timing information of each trigger to correlate events
\item Provide hit information of each hodoscope to determine the trajectory of incoming muons
\item Provide the charge information of each PMT output signal to determine the energy deposited inside the ND
\end{itemize}

\begin{figure*}[htb]
\begin{center}
\includegraphics[scale=0.43]{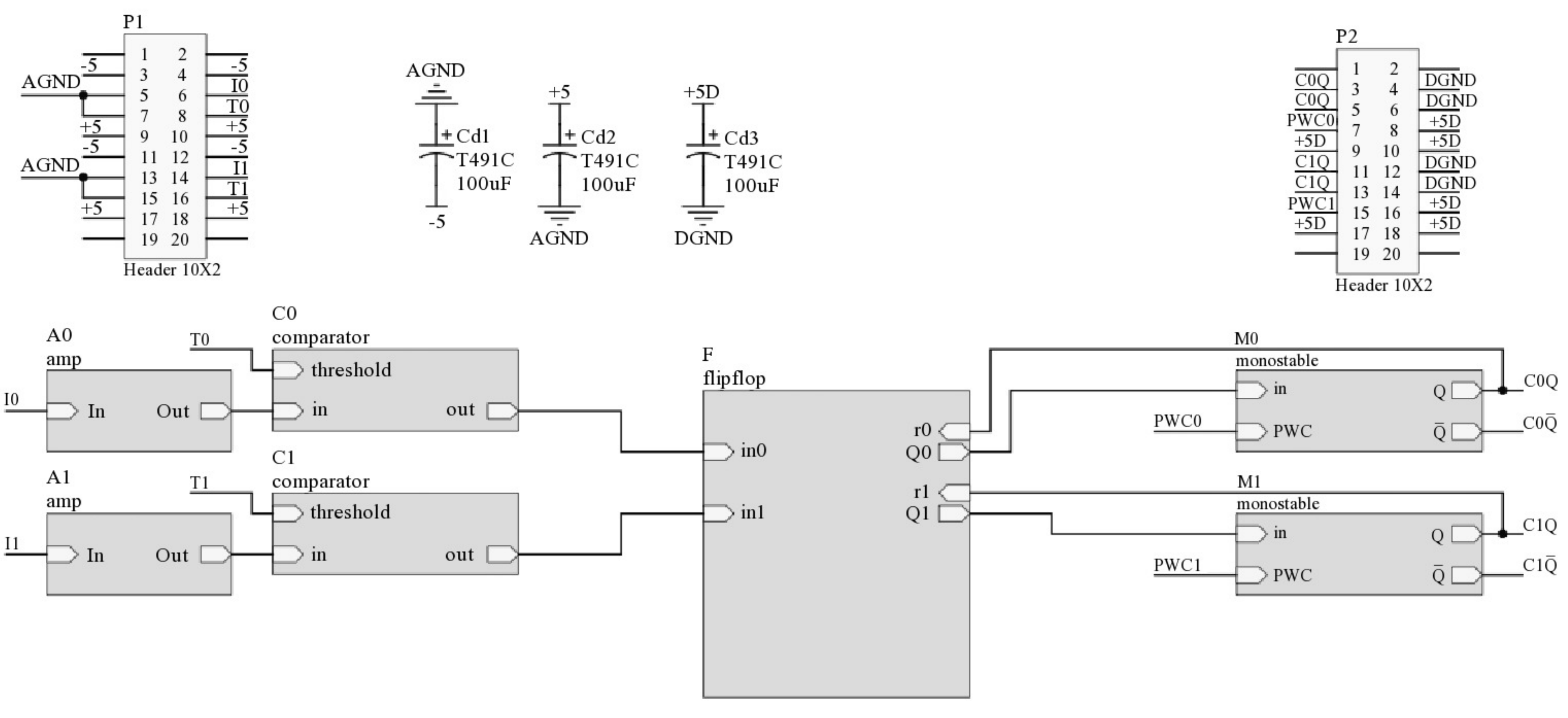}
\caption{ Schematic diagram of the muon tracker front-end daughter board (overview).}
\label{fig:14_mtfe_overview}
\end{center}
\end{figure*}

\begin{figure*}[htb]
\begin{center}
\includegraphics[scale=0.43]{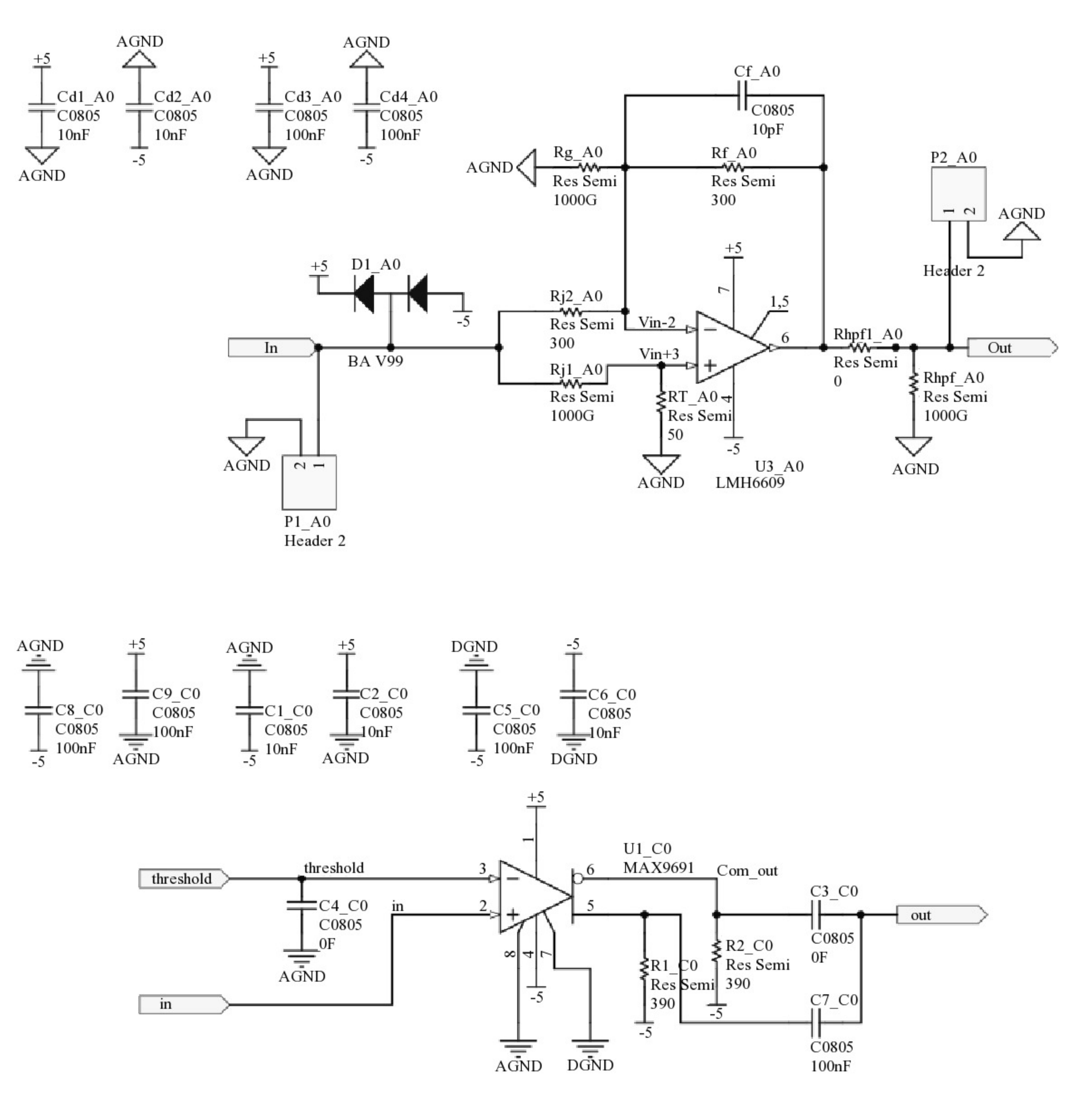}
\caption{ Schematic diagrams of the amplifier (upper) and comparator (lower) on the muon tracker front-end daughter board.} 
\label{fig:15_amp_comp}
\end{center}
\end{figure*}

\begin{figure*}[tb]
\begin{center}
\includegraphics[scale=0.43]{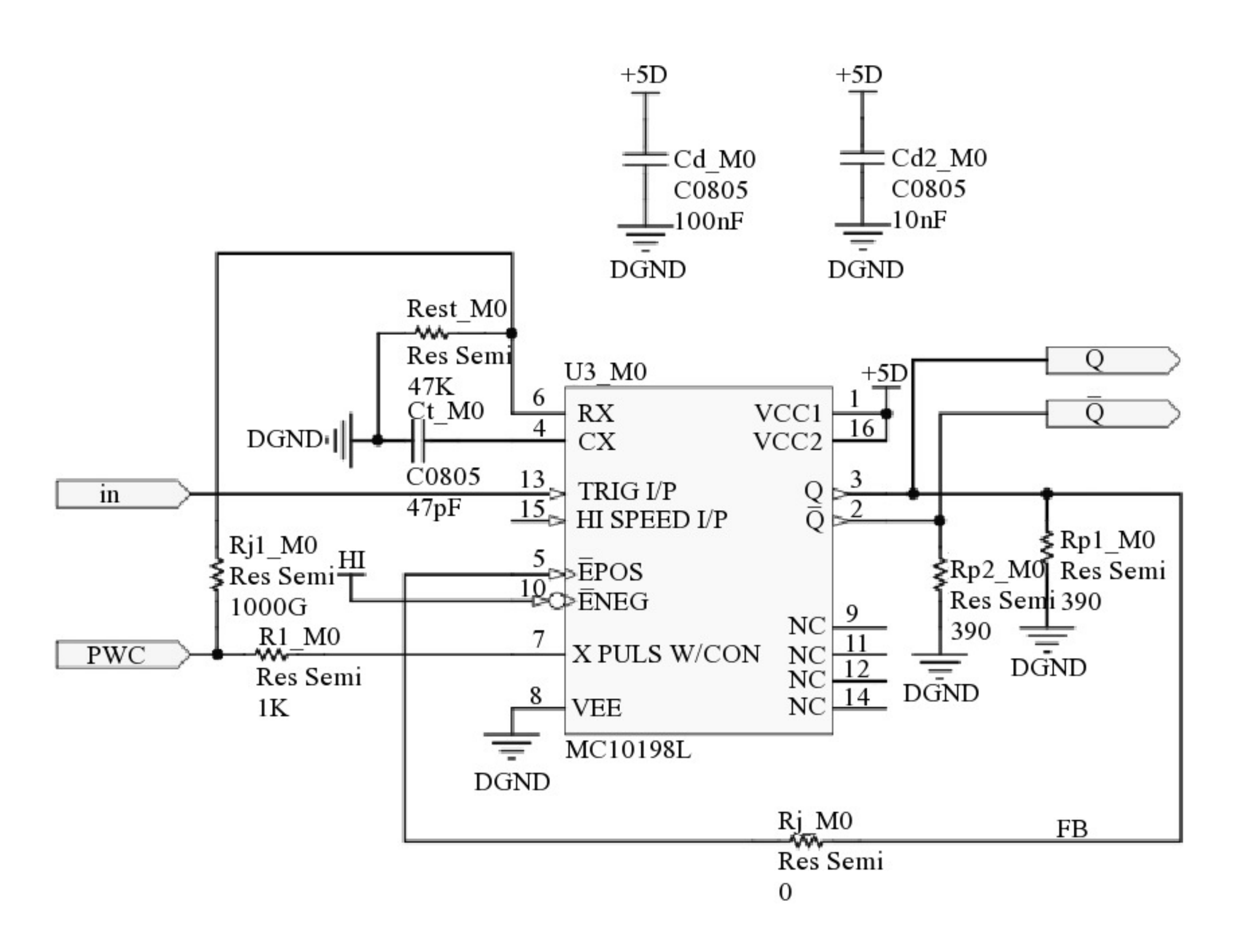}
\caption{ Schematic diagram of the monostable circuit on the muon tracker front-end daughter board.}
\label{fig:16_monostable}
\end{center}
\end{figure*}

The FEE for the MT consists of ten custom-designed 6U VME eight-channel discriminator boards, and a coincidence and pattern register module. Each discriminator board employs a modular design, with a mother board housing four daughter boards. The mother board provides power filtering, input and output connectors, and four 10-bit digital-to-analog converters (DACs) for setting the threshold and output pulse width remotely. Each daughter board has two input channels (Fig.~\ref{fig:14_mtfe_overview}), each of which has its own amplifier, comparator (Fig.~\ref{fig:15_amp_comp}) and a monostable circuit (Fig.~\ref{fig:16_monostable}) at the last stage. A typical value of the threshold is -35 mV. An ECL pulse of 100-ns wide is generated when the threshold is crossed.

The coincidence and pattern register for the MT are processed by using a CAEN VME V1495 module with on-board Field-Programmable Gate Array (FPGA) running at a clock of 100 MHz. This module receives the ECL signals from every MT FEE and processes the signals according to a preset trigger condition. Channels with signals can be masked individually at the input. The hit pattern of the MT is constructed inside the FPGA from the ECL signals and a preset mapping. If the hit pattern satisfies the trigger condition, that event is then latched and passed to an event builder inside the FPGA. The event builder adds a header to the event. This serves as a unique identifier for the start of the event and a redundant trigger state for cross-checking. Up to 500 events can be stored in the FIFO implemented in the FPGA.

\begin{figure*}
\begin{center}
\includegraphics[scale=0.55]{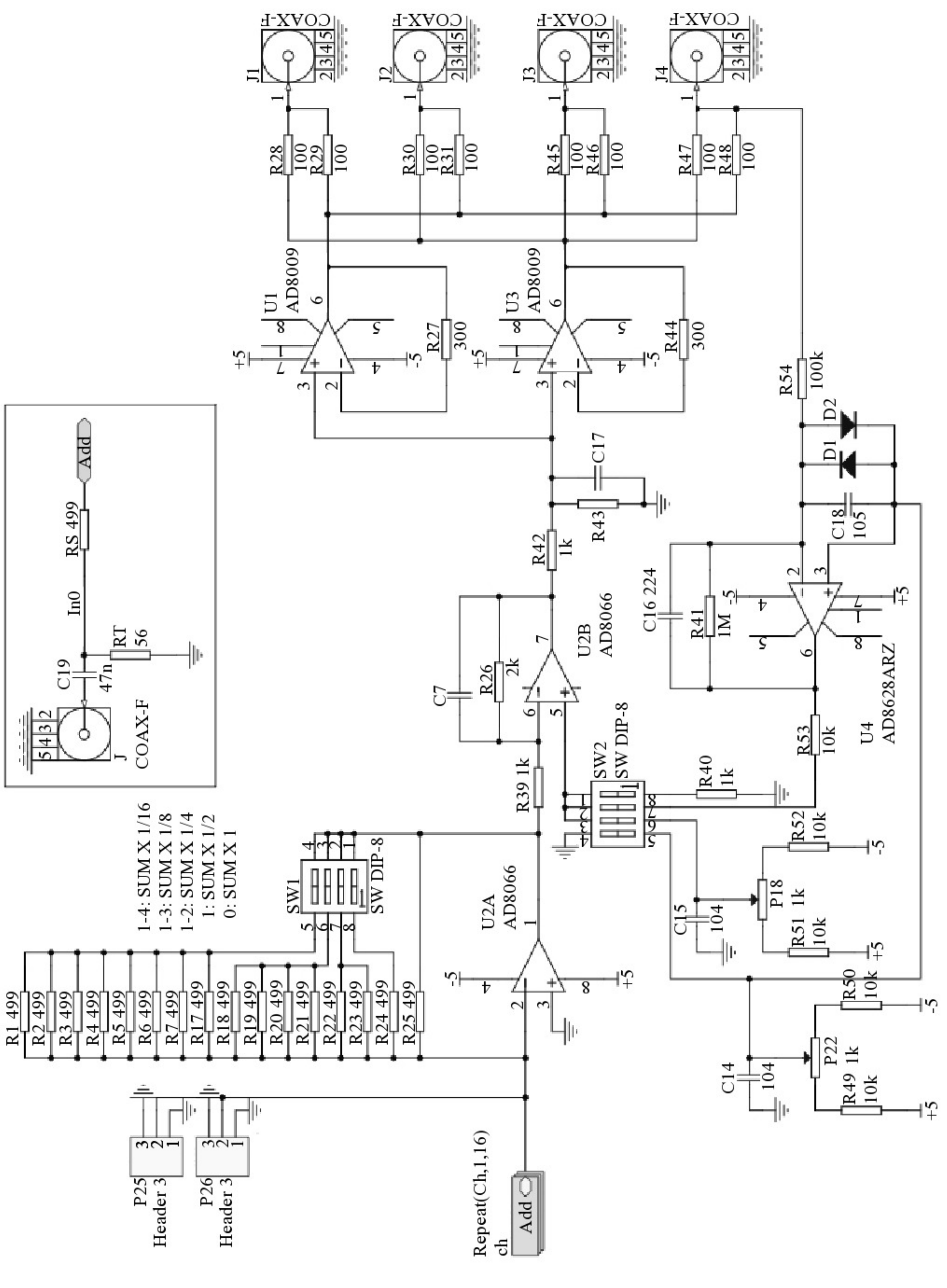}
\caption{ Schematic diagram of the energy-sum trigger module.}
\label{fig:17_esum}
\end{center}
\end{figure*}

For the ND, each PMT signal is duplicated with a CAEN V925 linear fan-in/fan-out module. One copy goes directly into a 12-bit QDC (CAEN V792N) for charge measurement. The other two copies go into a CAEN V895 leading-edge discriminator and an analog energy-sum (ESUM) module (Fig.~\ref{fig:17_esum}) respectively. The discriminator threshold of each channel is set in 1-mV-steps via the CAEN V1718 VME interface. An N-hit trigger will be generated if the signals received by the majority of PMTs exceed the threshold. The output of the ESUM module goes into the discriminator channel of the CAEN V925 that determines the energy threshold of the ESUM trigger. The logic signals of the N-hit, ESUM, and LED trigger from the ND calibration system are input to the CAEN V1495 module which serves as the Master Trigger Board (MTB) for the final trigger decision. A CAEN A395D I/O mezzanine board for handling the input of the logic signals is mounted on the MTB. 

The FPGA firmware of the MTB includes a trigger forming logic for the ND. An optional periodic trigger of the ND can be generated inside the MTB to monitor the pedestals and the background. The MTB also accepts the busy signals from the DAQ sub-systems of the MT and the ND respectively. Busy signals are generated during event building, charge conversion or when the event buffer is full. Thus a busy signal represents the presence of an accepted trigger. The MTB time-stamps the falling and rising edges of the busy signals with 10-ns resolution and records the corresponding event type (MT or ND). Events are then correlated with the time-stamps in the off-line analysis. Width of the busy signal is used to determine the dead-time of the DAQ.

\subsubsection{Triggers}\label{sec:trigger}
The passage of an energetic charged particle is identified by the temporal coincidence of signals from different hodoscope planes. It is flexible to form MT triggers with various coincidence combinations of the top (T), middle (M) and bottom (B) hodoscope layers. The 2-out-of-3 coincidence $H_{2/3}$ of either the x- or y-oriented hodoscope planes is defined as: 

\begin{equation}
H_{2/3} = T \bullet M + T \bullet B + M \bullet B
\end{equation}
where the outputs of the ten hodoscopes in each plane are logically OR-ed: 

\begin{align}
T &= \sum_{i=0}^{9} T_{i} \\
M &= \sum_{i=0}^{9} M_{i} \\
B &= \sum_{i=0}^{9} B_{i}
\end{align}
Similarly, the 3-out-of-3 coincidence $H_{3/3}$ for either the x- or y-plane
is defined as: 

\begin{equation}
H_{3/3} = T \bullet M \bullet B
\end{equation}
In order to reconstruct a muon track, at least two coordinates are required in both x- and y-direction. Therefore, ``2/3-x and 2/3-y'' ($H_{2/3-X} \bullet H_{2/3-Y}$) is required for muon track reconstructions. The trigger logic is implemented inside the FPGA for the MT.

For the ND, a N-hit (multiplicity) trigger is used as the primary trigger instead of an energy-sum (ESUM) trigger. The N-hit threshold can be set via the VME bus from 1 to 16 out of the total 16 PMTs. Triggers of the ND are accepted by the MTB only when the QDC is not busy, and when the buffers of the QDC and the MTB are not full. These criteria ensure the number of events registered by the QDC and the MTB to be identical. A trigger time window following an event from the MT can be imposed on the ND to reduce background events in the muon-induced neutron measurements. The trigger source (N-hit, ESUM, LED, periodic, or any combinations of the above) of each event is recorded in the MTB along with the time-stamp to facilitate event selections in off-line analysis.

\subsubsection{Data acquisition}\label{sec:acq} 
An open source data acquisition system called MIDAS (Maximum Integrated Data Acquisition System) \cite{bib:ritt-midas} is used as a skeleton of the DAQ firmwares. The system consists of a library and several applications which can run under major operating systems and can be ported easily to others.

The MIDAS library is written in the C programming language. The library contains routines for buffer management, a message system, a history system and an on-line database (ODB) \cite{bib:frlez}. The buffers between the producers and the consumers are FIFOs. The data transfer rate between a producer and a consumer over a standard 100BASE-TX Ethernet is on the order of 10 MB/s and can be higher if both run on the same computer. The history system is used to store slow control data and periodic events, which include event rate, high-voltage values, environmental variables or any data fields defined in the ODB, and produce time series plots. The on-line database is a central data storage for all relevant experiment variables such as run status, run information, front-end parameters, slow control variables, logging channel information and calibration constants. The ODB can be viewed and changed locally by using ODBEdit or remotely through a Web interface which is served by the MIDAS HTTP server. The password-protected Web interface provides a status overview of the experiment. Data taking can be controlled remotely through any internet browsers.

The implementation of MIDAS has one front-end computer connected to two VME crates, a CAEN SY1527LC high-voltage system, an environmental temperature/humidity sensor, a detector temperature sensor, and the calibration box for the ND. The VME crates and the environmental sensor are connected to the computer via high-speed USB 2.0. The detector temperature sensor and the calibration box are connected to the computer via RS232. The front-end computer communicates with the SY1527LC power supply via TCP/IP. Three MIDAS front-end programs are written and run on the same LINUX-based computer simultaneously. They consist of two parts: a system part which is linked to the MIDAS library for writing events into buffers and accessing ODB, and a user part which actually performs experiment- and hardware-specific data acquisition and control. The first one (trigger front-end) uses a polling scheme to read event responses of the MT and ND. The second one (slow control front-end) controls the high-voltage and measures temperature and humidity. The third one (calibration front-end) controls the calibration system of ND.

The front-end computer connects to the front-end electronics via high-speed USB 2.0. High-speed USB 2.0 hosts schedule transactions within microframes of 125 $\mu$s \cite{bib:axelson}. This makes an event-by-event polling of trigger events from the actual hardware inefficient. To accomplish a high event throughput using the polling scheme with USB 2.0, separate local buffers are maintained within the trigger front-end program. These local buffers are FIFOs and are managed by the user part of the program. During the polling cycle, if the local buffers are empty, the entire content in the buffer of the MTB is copied to a local buffer using one USB cycle. The trigger front-end program searches the MTB local buffer and counts the number of MT- and ND-tagged events respectively. Then the program copies events from the buffers of the corresponding modules (V1495 FPGA for MT-tagged events and V792N QDC for ND-tagged events) to their local buffers using one or two more USB cycles according to the number of tagged events in the MTB. If the local buffers are not empty during the polling cycle, a data-ready signal is issued and a single event is read from the local buffers. Data are read from the hardwares again when all the local buffers are empty.

The front-end computer connects to the back-end computer through a 100BASE-TX Ethernet. The back-end computer running under LINUX stores data to disks, performs on-line data analysis and hosts the MIDAS Web interface. Data files are written to a 500 GB disk in the back-end computer. Data can be transferred using the SCP protocol from the Aberdeen Tunnel laboratory to a 6-TB RAID 5 disk array in the University of Hong Kong for archive. To reduce the loading on the limited network bandwidth, data are also transferred by using external hard disk drives during access to the Aberdeen Tunnel laboratory. 

In the Aberdeen Tunnel experiment three basic on-line run types are defined: 
\begin{enumerate}
\item pedestal run;
\item calibration run;
\item physics run.
\end{enumerate}
The pedestal runs measure and update the QDC pedestal values for all channels of the ND. In a pedestal run, the N-hit and ESUM triggers of the ND and the trigger of the MT are disabled. Only the periodic trigger of 500 Hz will be sent to the ND to measure the QDC pedestal values. Pedestal subtraction and software gain correction routines in the on-line analysis are disabled and the raw pedestal values are filled to histograms. At the end of the pedestal run, the histograms are automatically fitted with Gaussian functions to obtain the mean pedestal values for all detector channels. The new pedestal values are updated to the ODB. 

The calibration runs acquire data for calculating the calibration constants of the ND in off-line analysis. In a calibration run, the trigger of the MT is disabled, and the N-hit trigger of the ND with N being set to 16 is enabled, with a discriminator threshold of -11 mV for each channel. This threshold enables the ND to be triggered by relatively low-energy gamma-rays such as the 0.66 MeV ones from $^{137}$Cs. A calibration run is often carried in two parts. The first one is a background run with no source in the ND. The trigger rate of a background run is typically about 5 kHz. After the background run, the calibration source is deployed to the designated position for a data run. When a calibration source is deployed at the ND center, the trigger rate is about 15 kHz for $^{60}$Co, and about 7 kHz for $^{137}$Cs.

During physics runs, the physics mode acquires events from the MT and ND. The type of events (MT or ND) is tagged and is stored together with the event. Data are calibrated on-line and are monitored through an on-line histogram manipulation tool Roody \cite{bib:amaudruz}. Muon tracks and the visible energy deposited in the ND are reconstructed. The reconstructed events can be visualized on-line using AbtViz as described in Section \ref{sec:vis}. A typical trigger rate of the MT is 0.013 Hz.

\subsubsection{Data model}
Events are stored in data files using the MIDAS binary format \cite{bib:ritt-fedata}. For this experiment, besides the default event headers, several data banks are defined in the trigger events for storing the data from the MT and ND; these includes a raw ADC bank, a raw muon hit pattern bank, and a calibrated ADC bank. The MIDAS logger compresses the data streams in the GNU-zipped format to reduce the size of the data files by about 50\%, while it takes about 20\% more CPU time \cite{bib:frlez}. The average size of a compressed event is 27 kB. The raw data files are processed event-by-event in the MIDAS analyzer through user-specific modules, with analyzer parameters and calibration constants stored in the ODB. Then the calibrated data and reconstructed events are written into a ROOT file with a TTree object container provided by a ROOT class. The data can be accessed through more than thirty ROOT classes with the interactive C++ interpreter CINT embedded in ROOT. With the ROOT data files, the users can analyze the raw data, calibrated data, reconstructed events, detector modeling, data calibration, energy calibration, event reconstruction, and run information.

\subsection{Event visualization}\label{sec:vis}

\begin{figure}
\begin{center}
\includegraphics[scale=0.2]{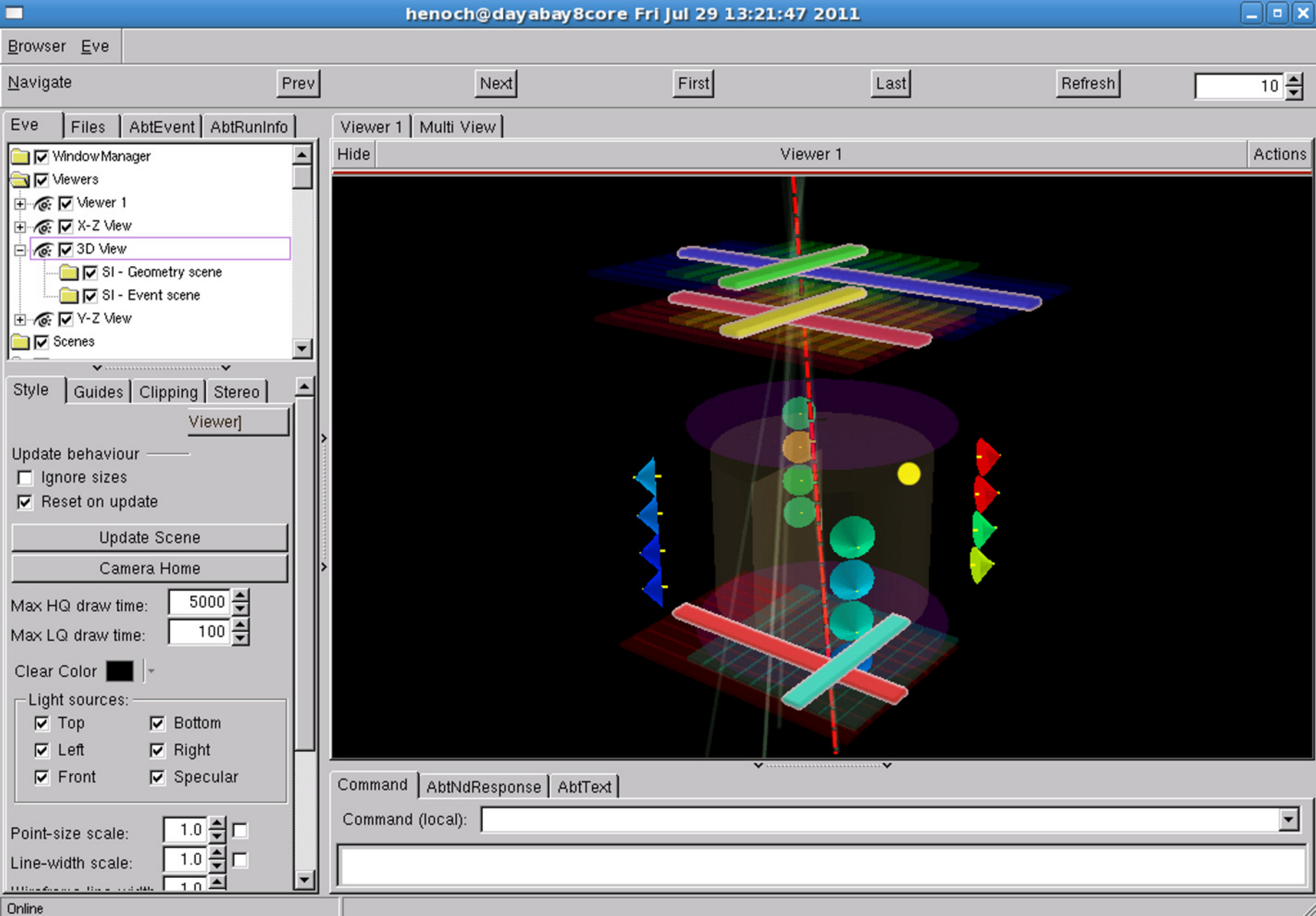}
\end{center}
\caption{GUI controls and OpenGL display of an event in AbtViz}
\label{fig:18_abt_viz}
\end{figure}

The Aberdeen Event Visualization (AbtViz) is developed using the Event Visualization Environment \cite{bib:eve}, a high-level visualization library using ROOT's \cite{bib:root} data-processing, GUI and OpenGL interfaces that emerged from the development of the event display of the ALICE experiment at the Large Hadron Collider. The program is written in the iPython \cite{bib:ipython} environment to take advantage of the enhanced interactivity for visual debugging the simulation and reconstruction algorithms.

The geometries are exported directly from a GEANT4 simulation \cite{bib:g4} using the Virtual Geometry Model \cite{bib:vgm} package with TGeo objects, which is accessible by the ROOT-based visualization. As shown in Fig.~\ref{fig:18_abt_viz}, the MT hodoscopes are represented as boxes, ND PMTs as cones, reconstructed muon tracks as lines, and the reconstructed neutron vertex as a sphere. The detector responses are visualized by highlighting the triggered hodoscopes and each individual PMT. Their color is determined by mapping its ADC signal value to the RGBAPalette class.

Event-data are stored as ROOT TObject. They are sent via pika 0.9.5 \cite{bib:pika}, a pure-Python implementation of the AMQP 0-9-1 protocol \cite{bib:amqp}, directly from the Aberdeen Tunnel DAQ machine to a RabbitMQ \cite{bib:rabbitmq} server situated at the Chinese University of Hong Kong. By subscribing remote AbtViz to the message queue, the event-data are received in real-time for on-line analysis.

\section{Performance of Apparatus}\label{sec:perform}
\subsection{Muon tracker}
\subsubsection{Hodoscope efficiency}
\begin{figure}[h]
\centering
\includegraphics[width=0.5\textwidth]{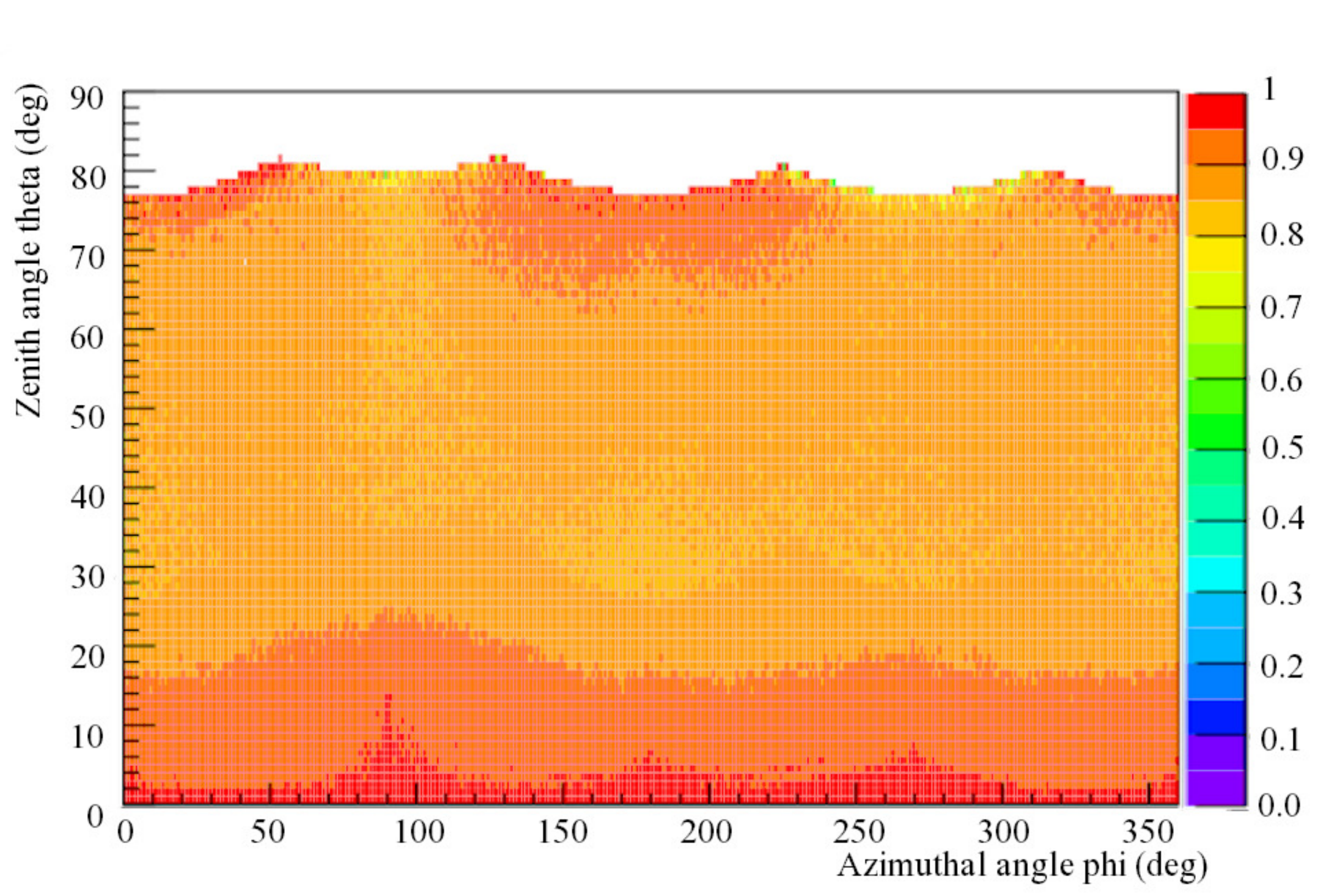}
\caption{Efficiency of the MT plastic-scintillator hodoscopes  as a function of the zenith and azimuthal angles.}
\label{fig:19_muon_ang_eff}
\end{figure}

The efficiency of a hodoscope of the MT is determined by using muons acquired with the MT. Only muons that go almost vertically through the MT are selected. In these events, the overlapping area of all the fired hodoscopes forms a rectangle conveniently defined by the widths of the plastic scintillators. In the analysis, for a particular section $(i,j)$ of a hodoscope in plane $k$ amongst the $m$ planes, its efficiency is determined by the number of 6-fold coincidences ($N_{i,j; 6-fold}$) and 5-fold coincidences that exclude the hodoscope of interest ($N_{i,j; m \ne k}$) as

\begin{equation}
\label{eq:mt_eff}
\frac{N_{i,j; m \ne k}}{N_{i,j; 6-fold}} = \frac{\prod_{m=1; m \ne k}^{6} \epsilon(i,j;m) R_{\mu}t}{\prod_{m=1}^{6} \epsilon(i,j;m) R_{\mu}t} 
= \epsilon(i,j;k)
\end{equation}
where $R_{\mu}$ is the muon rate and $t$ is the measurement time. Following the configuration shown in Fig.~\ref{fig:4_mt_1}, the hodoscopes aligned in the east-west direction are divided into $j$ sections, those running along north-south are partitioned into $i$ sections. The efficiency of each section is obtained with Eq. \ref{eq:mt_eff}. The average efficiency of the hodoscopes on the top or in the middle layer is 95\% $\pm$ 4\%. For the bottom layer, the efficiency of the hodoscopes was determined by sandwiching each of them with two reference hodoscopes. The efficiency of the individual hodoscope in the bottom layer is about 96\% $\pm$ 4\%, with an exception of four 1.5-m-long hodoscopes of which the efficiency does not depend on the hit position along the length of the plastic scintillator. Fig.~\ref{fig:19_muon_ang_eff} shows the efficiency of the whole MT after integrating over the solid angle subtended by an area covered by the 2-out-of-3 coincidence of the hodoscopes running in the east-west direction and the 2-out-of-3 coincidence of the hodoscopes along north-south ($H_{2/3-X} \bullet H_{2/3-Y}$). 

\subsection{Neutron detector}
\subsubsection{Gamma-ray background in the ND} \label{sec:gamma_background}

GEANT4 toolkit \cite{bib:g4} is used to estimate the gamma-ray background coming from rock and detector materials. In the simulation, gamma-rays from rock are generated based on the results obtained with a HPGe detector as discussed in Section \ref{sec:lab_env}. In addition, radioactivities of a sample of the steel frame and a Hamamatsu R1408 PMT were measured with an Ortec GEM 35S HPGe detector placed inside a 10-cm-thick low-background lead shield (Canberra 767). The results are summarized in Table \ref{table:gamma_conc_pmt_ss}. Using detailed descriptions of the detector geometry and components of the ND, gamma-rays from the steel frame of the MT, the sixteen Hamamatsu R1408 PMTs, and the two retroreflectors for monitoring the optical transmittance of the mineral oil are also simulated. 

\begin{table}[ht]
\begin{center}
\begin{tabular}{c c c c}
\hline
Isotope & R1408 & MT frame & Retroreflector\\
\hline\hline
$^{238}$U & 118$\pm$5 & 7$\pm$6 & 1.7$\pm$ 0.2 \\
$^{232}$Th & 36$\pm$9 & 13$\pm$10 & 1.1$\pm$0.3 \\
$^{40}$K & 2820$\pm$80 & 41$\pm$6 & 520$\pm$2 \\
\hline
Unit weight & 0.68 kg & 1 kg & 1 kg \\
\hline
\end{tabular}
\caption{ Activity (Bq/kg) of $^{238}$U, $^{232}$Th and $^{40}$K in different detector materials.}
\label{table:gamma_conc_pmt_ss}
\end{center}
\end{table}

\begin{figure}[ht]
\begin{center}
\includegraphics[scale=0.305]{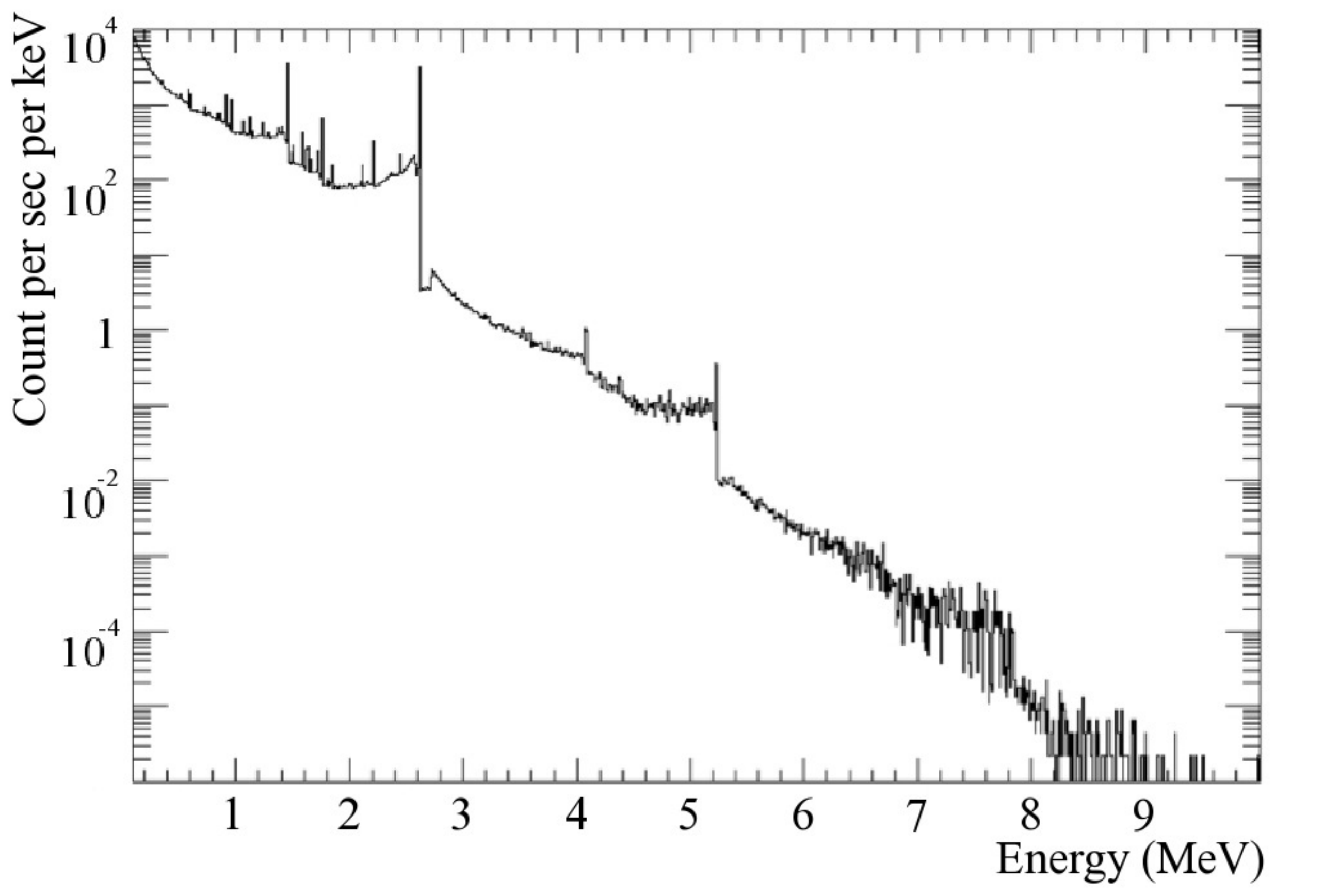}
\caption{Simulated gamma-ray coincidence background in a time window of 50 $\mu$s for the ND in the Aberdeen Tunnel laboratory.}
\label{fig:20_gamma_nd}
\end{center}
\end{figure}

A time window of 50 $\mu$s is implemented in the code to simulate coincidence of gamma-rays. Fig.~\ref{fig:20_gamma_nd} is the simulated energy spectrum of the coincident gamma-rays that enter the ND. With a threshold of 5.3 MeV, the coincident gamma-ray background can be suppressed to the level of 0.6 Hz, down by more than 3 orders of magnitude. In fact, the observed background rate is only 0.24 Hz $\pm$ 0.04 Hz.

\subsection{Energy calibration of the neutron detector}\label{subsubsec:en_scale}
\begin{figure}[t]
\begin{center}
\includegraphics[width=0.45\textwidth]{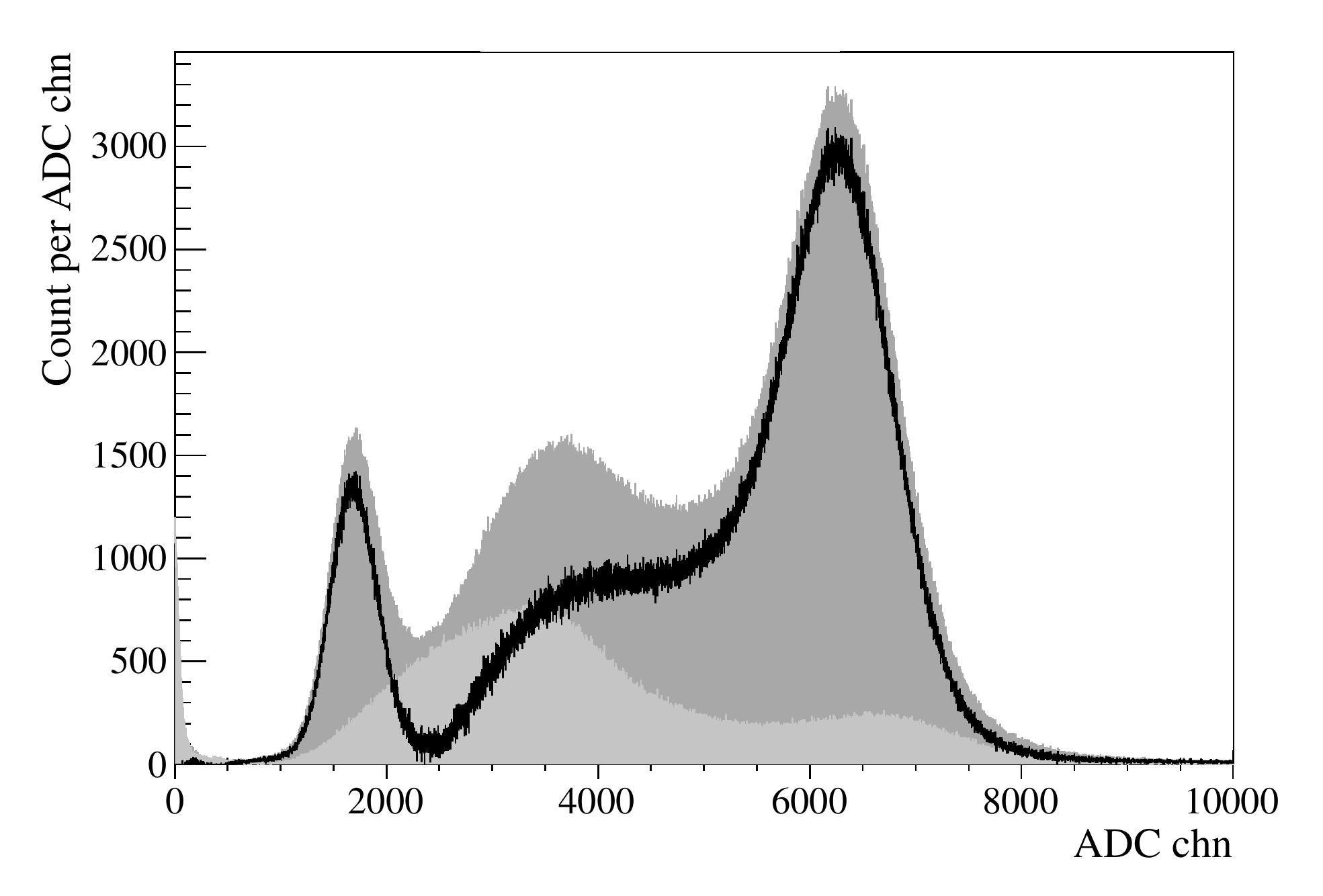}
\caption{ADC distribution obtained with and without $^{60}$Co and $^{137}$Cs sources. The dark gray spectrum is generated by the $^{60}$Co and $^{137}$Cs positioned at the center of the Gd-LS.  The light gray spectrum is due to background gamma-rays detected by the ND. The black histogram is the background-subtracted distribution of the calibration sources. The peak on the left corresponds to the 0.66-MeV gamma-ray of $^{137}$Cs. The peak on the right is the total energy deposited by the (1.17+1.33)-MeV gamma-rays of $^{60}$Co.}
\label{fig:21_cal_co_cs}
\end{center}
\end{figure}

\begin{figure}[t]
\begin{center}
\includegraphics[width=0.45\textwidth]{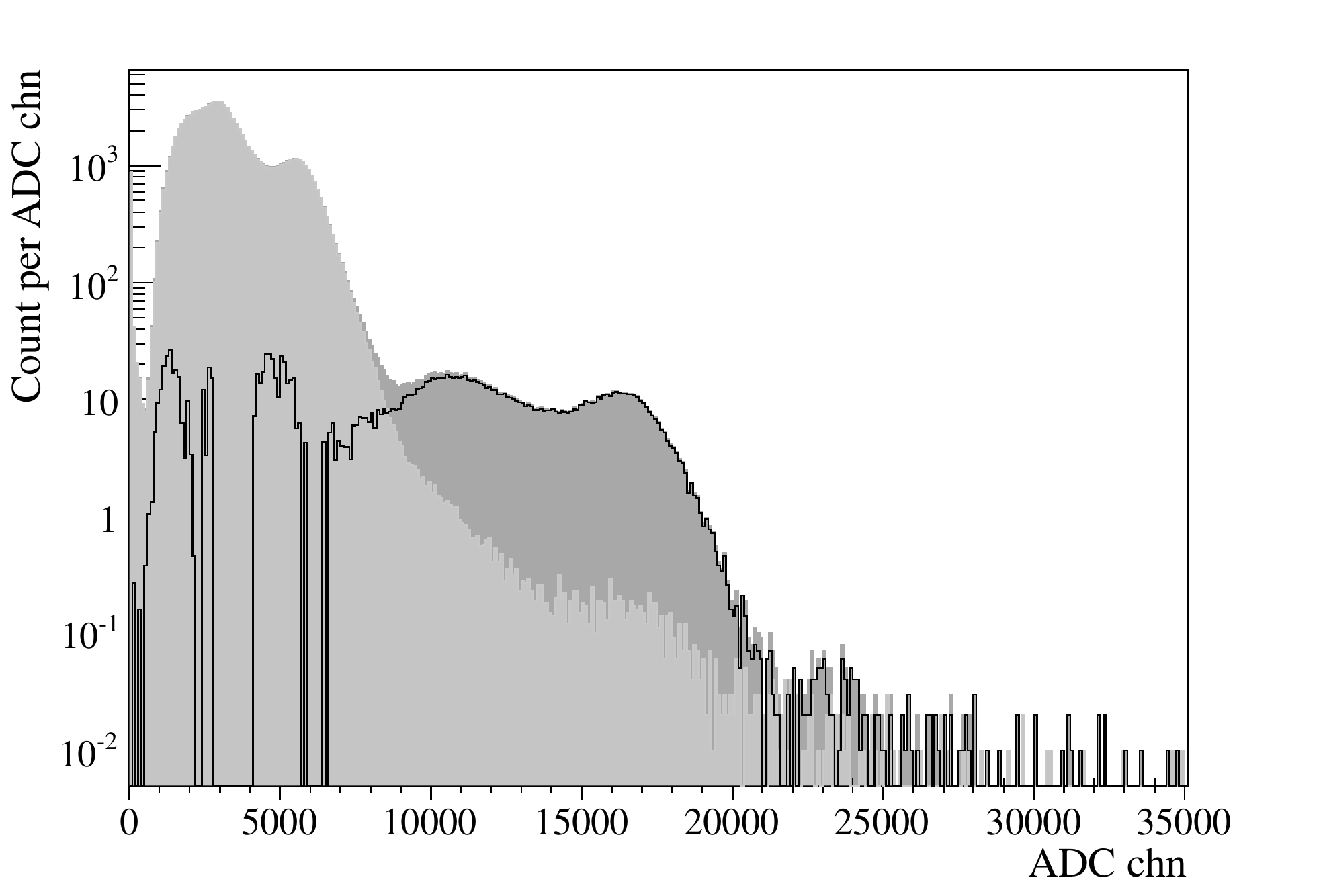}
\caption{ADC distribution obtained with and without the Am-Be source. The dark gray spectrum is for Am-Be at the center of Gd-LS volume. The light gray spectrum is the background gamma-rays seen by the detector. The black histogram is background-subtracted spectrum of the Am-Be source. The peak near channel 16,000 is the total energy of about 8 MeV released by the gamma-rays when a Gd nucleus captures a neutron.}
\label{fig:22_cal_ambe}
\end{center}
\end{figure}

Fig.~\ref{fig:21_cal_co_cs} shows the energy spectra before and after background subtraction 
obtained with the $^{137}$Cs and $^{60}$Co sources. A typical energy distribution for the Am-Be source is illustrated in Fig.~\ref{fig:22_cal_ambe}. The peaks with energies less than 3 MeV are due to gamma-rays coming from natural radioactivity in the vicinity of the ND whereas the peaks greater than 3 MeV are related to a sequence of events leading to neutron capture on Gd. 

\begin{figure}[h]
\begin{center}
\includegraphics[width=0.48\textwidth]{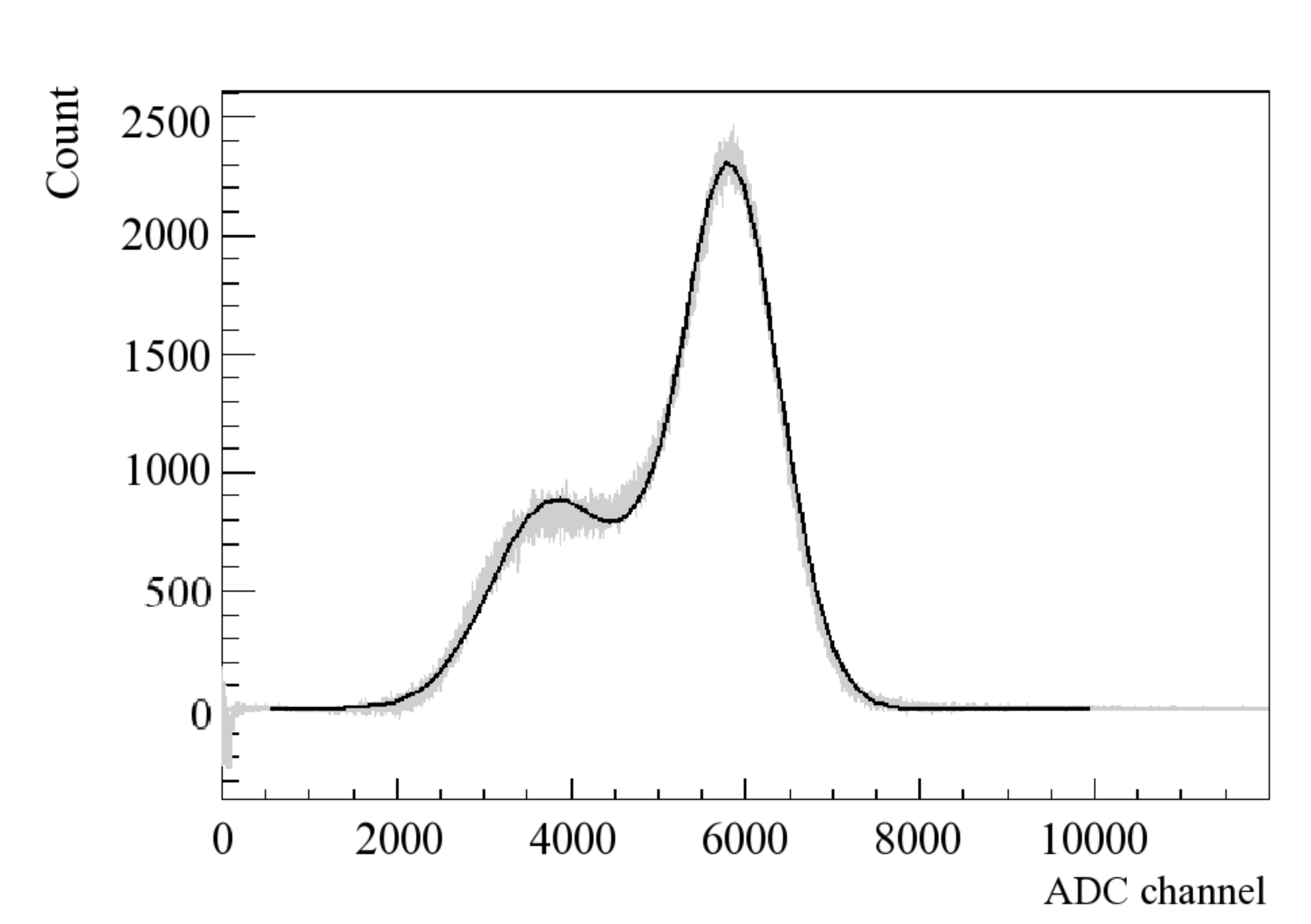}
\caption{ $^{60}$Co spectrum fit by a Crystal Ball function and a Gaussian. }
\label{fig:23_crystal_ball}
\end{center}
\end{figure}

In the off-line analysis, after subtracting the background obtained from the background run, the peaks in the ADC spectra of the $^{137}$Cs and Am-Be sources are fitted to Gaussian distributions. For the $^{60}$Co source, the distribution is fitted with a Crystal Ball function \cite{bib:crystal_ball} plus a Gaussian in the low-energy side of the peak (Fig.~\ref{fig:23_crystal_ball}). The Crystal Ball function is given by:

\begin{equation}
f(x; \alpha, n, \bar{x}, \sigma) = N
\begin{cases}
\rm{exp}( -\frac{ (x-\bar{x})^2 }{ 2 \sigma^{2} } ), & \text{if}\ \frac{x-\bar{x}}{\sigma} > -\alpha \\
      A(B-\frac{x-\bar{x}}{\sigma})^{-n} & \text{if}\ \frac{x-\bar{x}}{\sigma} \leq -\alpha
\end{cases} 
\end{equation}
where
\begin{align}
A & = \bigg( \frac{n}{ \left| \alpha \right| } \bigg)^{n} \centerdot \rm{exp} \bigg( - \frac{\left| \alpha \right|^{2}}{2} \bigg) \\
B & = \frac{n}{ \left| \alpha \right| } - \left| \alpha \right| 
\end{align}
and $N$ is the normalization factor, $n$, $\alpha$, $\bar{x}$ and $\sigma$ are the fitting parameters. The Gaussian component of the Crystal Ball function is used to model the peak resulting from the two gamma-rays with similar energies in the $^{60}$Co spectrum. The additional Gaussian is used to take care of the contribution of one of the gamma-rays from $^{60}$Co when the other one cannot deposit all of its energy in the fiducial volume of the ND. The average calibration constants determined with the radioactive sources from April 2011 to November 2012 are tabulated in Table \ref{table:average_en_scale}. The results are consistent with each other within two standard deviations.

\begin{table}[h]
\begin{center}
\begin{tabular}{ c c}
\hline
Source 	& 	Average energy scale	\\
		&	(ADC counts/MeV)		\\
\hline \hline
$^{137}$Cs	&	2300 $\pm$ 34		\\
$^{60}$Co	&	2306 $\pm$ 35		\\
Am-Be neutron	&	2243 $\pm$ 20		\\
\hline
\end{tabular}
\caption{Average energy calibration constants determined with calibration sources located at the ND center.}
\label{table:average_en_scale}
\end{center}
\end{table}

\subsection{Monitoring energy scale of the neutron detector}
\begin{figure*}[ht]
\centering
\includegraphics[width=0.77\textwidth]{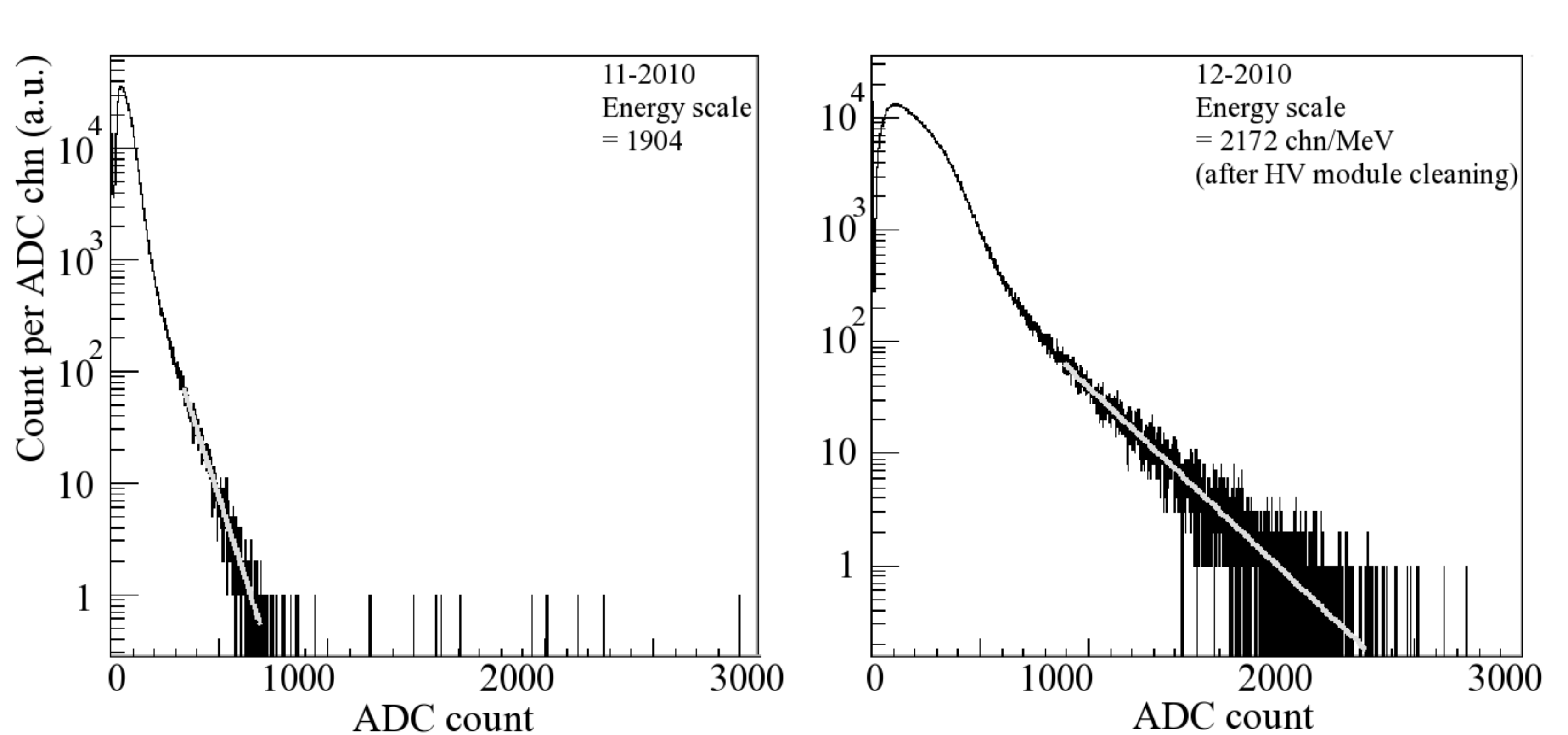}
\caption{Charge distributions of PMT NW4 (north-west corner, ring 4) before (left) and after (right) cleaning the HV module. The gray lines are the fitted exponential functions for determining the endpoints.}
\label{fig:24_PMT_NW4_end_point}
\end{figure*}

Energy calibration (Section \ref{subsubsec:en_scale}) is performed regularly by deploying  the calibration sources ($^{137}$Cs, $^{60}$Co, Am-Be) to the center of the ND. This serves as a regular check of the operational stability of the apparatus that includes the DAQ, optical quality of the liquids and PMT gain. The CAEN SY1527LC high-voltage power supply mainframe and the CAEN A1535SN high-voltage supply modules are found to be very sensitive to the dusty environment in the Aberdeen Tunnel laboratory. In about four months of operation, as demonstrated in Fig.~\ref{fig:24_PMT_NW4_end_point}, the high-voltage supplied to the PMTs can drop by as much as 16\% from the read-back values, resulting in a drop of PMT gain. Cleaning the high-voltage system with dry compressed air or perform a factory calibration can reduce the discrepancies.

\begin{figure*}[t]
\centering
\includegraphics[scale=0.6]{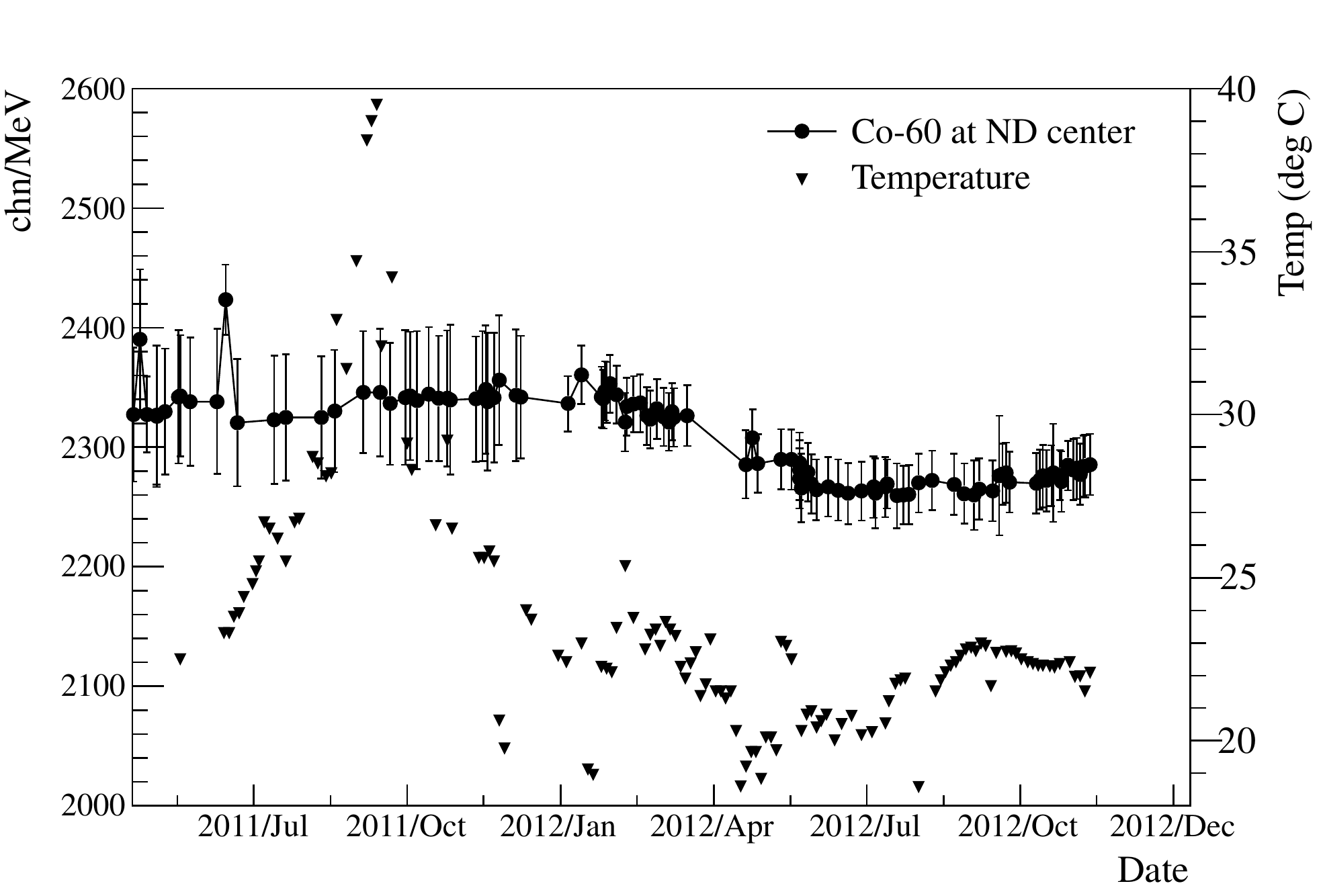}
\caption{Corrected calibration constant of the neutron detector for the $^{60}$Co source as a function of time. 
The variation of the temperature in the underground laboratory is also shown. The rise in temperature in August and September of 2011 was due to a failure of the air-conditioning unit.}
\label{fig:25_en_scale_zoom}
\end{figure*}

In order to compensate for the fluctuation in the detector response, a correction factor is introduced to each PMT based on its own output charge distributions. This factor is taken to be a ratio of the endpoint positions of the singles spectra of the same PMT. The average position of the endpoint obtained in the first month since the installation of the high-voltage system is taken as the reference. The fitting range used for determining the endpoint is between 0.4 MeV and 1 MeV, in which the spectral distortion due to a change in the trigger condition is not observed. Before determining the endpoints of the spectra using an exponential function, the distributions in the fitting range are normalized. The corrected energy scale of the ND as a function of time is plotted in Fig.~\ref{fig:25_en_scale_zoom}. Each PMT has its own correction factor. The overall effect is that the corrected energy scale is about 1.26 times of the value before correction. It is interesting to note that an increase in temperature from about 22$^{\circ}$C to 40$^{\circ}$C in the underground laboratory for about two months due to a failure of the air-conditioning unit did not degrade the performance of the apparatus, in particular, the Gd-LS.

\subsubsection{Response of neutron detector to gamma-ray sources} 
GEANT4 allows implementation of the detector response, in particular the PMT optical model from GLG4Sim. Response of the ND to the gamma-ray calibration sources at different positions has been simulated. Gamma-rays of the corresponding energies are generated isotropically inside the active volume of the calibration source. The following electromagnetic processes related to gamma-rays are included: ionization, bremsstrahlung, photoelectric effect, fluorescence, Rayleigh scattering, Compton scattering and pair production.  Besides, attenuation length of the Gd-LS, and optical properties of the ND such as the reflectivity of the top, bottom, and side reflectors is considered. The charge resolution and gain of individual PMT are also implemented in the simulation.

\begin{figure}[p]
\centering
\includegraphics[width=0.48\textwidth]{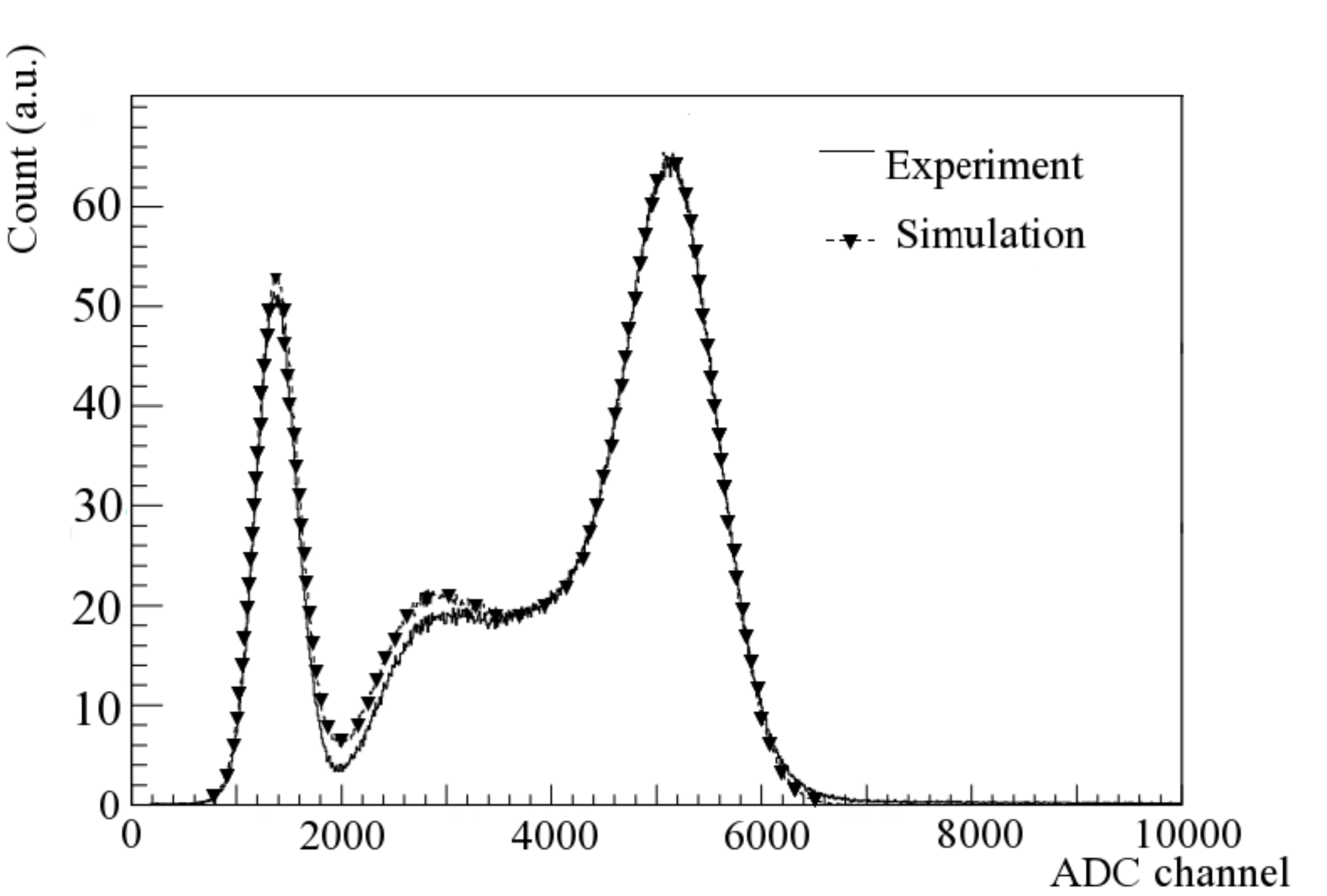}
\caption{Comparison of measured energy spectrum of $^{137}$Cs and $^{60}$Co with the simulated spectrum obtained with GEANT4.}
\label{fig:26_CTCenterCompare-BW}
\end{figure}

Fig.~\ref{fig:26_CTCenterCompare-BW} shows the simulated and observed energy spectra for the $^{137}$Cs and $^{60}$Co source placed at the center of the Gd-LS volume. There is a good agreement between the experimental and simulated results. From simulation, the energy resolution at the center of the ND at 0.66 MeV and 2.5 MeV are 13.5\% and 9.3\%, in agreement with the experimental measurements of 14.3\% $\pm$ 0.2\% and 9.5\% $\pm$ 0.2\% respectively. 

\subsubsection{Response of neutron detector to the Am-Be source} 
The Am-Be source (Fig.~\ref{fig:12_AmBeGeom}) used for calibration emits neutrons and gamma-rays through the following channels:
\begin{equation}
\label{eq:ambe_channels}
\begin{split}
^{241}\rm{Am} \rightarrow \ ^{237}\rm{Np} + \alpha  \\ 
n_b: \ ^{9}\rm{Be} +  \alpha \rightarrow&  \ ^{9}\rm{Be}^{*} + \alpha' \rightarrow n + \ ^{8}\rm{Be} +  \alpha' \\
n_0: \ ^{9}\rm{Be} +  \alpha \rightarrow& \ n + \ ^{12}\rm{C} \\
n_1: \ ^{9}\rm{Be} +  \alpha \rightarrow& \ n + \ ^{12}\rm{C}^{*} \ (4.4 \ MeV \ \gamma)  \\
n_2: \ ^{9}\rm{Be} +  \alpha \rightarrow& \ n + \ ^{12}\rm{C}^{*} \ (4.4 + 3.2 \ MeV \ \gamma s)
\end{split}
\end{equation}
 
The energy spectrum obtained with the Am-Be source in the ND is a combination of the energies deposited by the neutrons and gamma-rays through various interactions, including thermalization and
capture of neutron in the Gd-LS, and energy released in the active
volume by the gamma-rays from the decay of $^{12}$C$^{*}$ (for $n_{1}$ and $n_{2}$). In addition, the decay of $^{12}$C$^{*}$ has a short half-life of femtoseconds. The gamma-rays from the $^{12}$C$^{*}$ de-excitation are detected in coincidence with the signal due to neutron thermalization. Thus, it is necessary to implement the energy- and time-correlation of neutrons and associated gamma-rays of the Am-Be source in the simulation in order to reproduce the observed energy spectrum.

The energy spectrum of the neutrons emitted by an Am-Be source, and the neutron energy distribution of each channel in Eq.~\ref{eq:ambe_channels} have been reported in Refs. \cite{bib:Geiger1975}\cite{bib:Marxh1995}. These published results are used in the simulation. From the picked neutron energy, the corresponding production channel is identified, and the energy of the associated gamma-rays is generated accordingly. 

\begin{figure}
\centering
\includegraphics[width=0.5\textwidth]{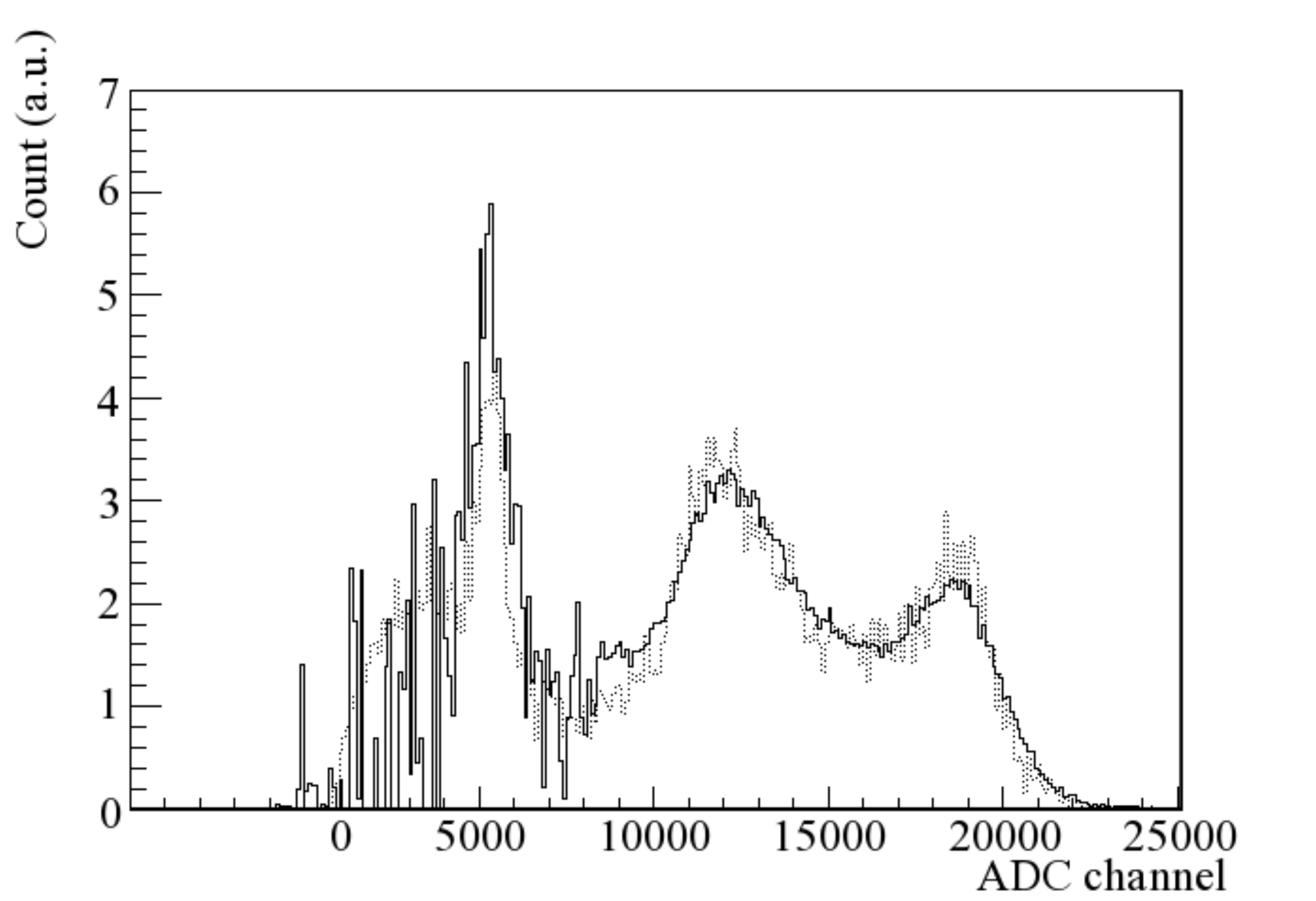}
\caption{Comparison of the observed (solid) and simulated (dotted) ADC spectra for the case with the Am-Be source deployed at the center of the neutron detector. The peaks around the ADC channels 18,000, 12,000, and 5,000 correspond to energies of 8 MeV, 5.3 MeV, and 2.2 MeV respectively.}
\label{fig:27_AmBeSpec_cfSim}
\end{figure}

The arrival times of the optical photons at the PMTs are recorded for simulating realistic temporal correlation of events. The ADC sum of the sixteen ND PMTs are plotted in Fig.~\ref{fig:27_AmBeSpec_cfSim}. In the figure, the peak near channel 5,000 corresponds to the
gamma-rays released from neutron capture on protons, whereas the distribution peaked near channel 18,000 is due to the gamma-rays generated from neutron capture on Gd nuclei in the Gd-LS. The broad distribution with a peak around channel 12,000 is the sum of energy deposited by the 4.4-MeV gamma-ray ($n_{1}$, $n_{2}$) and proton recoil during thermalization of the energetic neutrons from the Am-Be source. Again, the simulated energy distribution is consistent with the observed spectrum.

\subsubsection{Detection of spallation neutrons produced by cosmic-ray muons}
\begin{figure}
\centering
\includegraphics[width=0.48\textwidth]{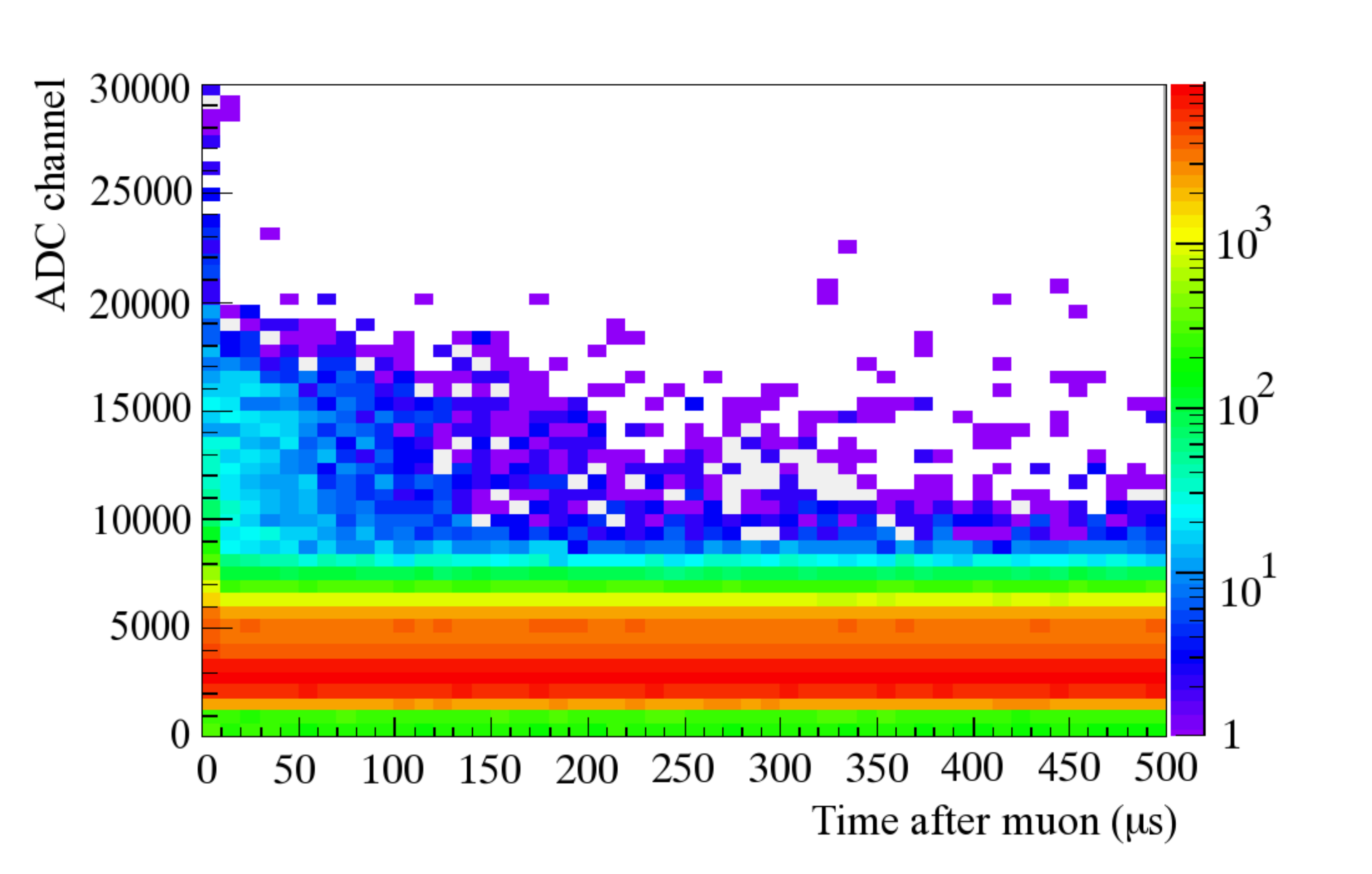}
\caption{ND visible energy (in ADC channels) versus ND trigger time after the last MT trigger. Neutron-capture events of muon-induced neutrons clustered around ADC channel 16,000,  from 0 $\mu$s to 200 $\mu$s.}
\label{fig:28_hNdCapTimeAdc_grayscale}
\end{figure}

To demonstrate the capability of the ND to observe spallation neutrons induced by cosmic-ray muons, the detected energy (in units of ADC channel) of events seen in the ND versus the first occurrence of the ND trigger after the last MT trigger is shown in Fig.~\ref{fig:28_hNdCapTimeAdc_grayscale}. The data depicted in the plot were collected in 128 days and without any event selection requirement imposed. Most of the events between ADC channels 0 and 10,000 are accidental gamma rays that are uncorrelated with the cosmic-ray muons. However, the events clustered around ADC channel 16,000, corresponding to an energy of 8 MeV, are correlated with the MT trigger between 0 $\mu$s and 200 $\mu$s. These are events due to neutrons captured by Gd in the Gd-LS, hence providing the evidence that muon-induced neutrons have been observed in the ND.  

\section{Conclusion}

We have successfully constructed a plastic-scintillator tracker and a neutron detector for studying spallation neutrons produced 
by cosmic-ray muons in the Aberdeen Tunnel laboratory in Hong Kong. The equipment has been in routine operation for about a year. The average efficiency of the scintillator hodoscopes is better than 95\%. The energy response of the neutron detector containing 650~kg of 0.06\%-Gd-LS has been studied and monitored with gamma-rays emitted by radioactive sources placed at the center of the detector. The capability of the neutron detector in detecting low-energy neutrons has been demonstrated with an Am-Be source. In general, the performance of the apparatus is consistent with expectation based on comparisons with simulation.

\section{Acknowledgement}
This work is partially supported by grants from the Research Grant Council of the Hong Kong Special Administrative Region, China (Project nos. HKU703307P, HKU704007P, CUHK 1/07C and CUHK3/CRF/10), University Development Fund of The University of Hong Kong, and the Office of Nuclear Physics, Office of High Energy Physics, Office of Science, US Department of Energy under the Contract no. DE-AC-02-05CH11231, as well as the National Science Council in Taiwan and MOE program for Research of Excellence at National Taiwan University and National Chiao-Tung University.

The authors would like to thank the Commissioner for Transport, The Government of the Hong Kong Special Administrative Region, for providing the underground facilities, and Serco Group plc, for their cooperation and support in the Aberdeen Tunnel. \\


\end{document}